\documentclass[preprint,12pt,authoryear]{elsarticle}
\usepackage{graphicx} 
\usepackage{amsmath}
\usepackage{rotating}
\usepackage{caption}
\usepackage{makecell}
\usepackage{subfigure}
\usepackage{comment}
\usepackage{bm}
\usepackage{amssymb}
\usepackage{scalerel}
\usepackage{esvect}
\usepackage{xcolor}
\usepackage{upgreek}
\usepackage[flushleft]{threeparttable}
\usepackage{booktabs}
\usepackage{soul}
\usepackage{array}
\usepackage{multirow}
\usepackage{longtable} 
\usepackage{ragged2e}
\usepackage{tabularx}  
\usepackage[colorlinks=true, allcolors=blue]{hyperref}

\journal{xxx}
\begin{document}
\begin{frontmatter}

\title{Machine Learning for Exoplanet Discovery: Validating TESS Candidates and Identifying Planets in the Habitable Zone}
\author[1]{Sarah Huang}
\ead{sarah.huang2009@outlook.com}

\author[2]{Chen Jiang\corref{cor1}}
\ead{jiangc@mps.mpg.de}
\cortext[cor1]{Corresponding author}

\affiliation[1]{organization={Shanghai American School, Puxi Campus},
           addressline={Jinfeng Road 258, Minhang District}, 
           city={Shanghai},
           postcode={201107}, 
           country={China}}
           
\affiliation[2]{organization={Max-Planck-Institut für Sonnensystemforschung},
            addressline={Justus-von-Liebig-Weg 3}, 
            city={Göttingen},
            postcode={37077}, 
            country={Germany}}

\date{April 2025}

\begin{abstract}
The high-precision photometry from NASA's Kepler and TESS missions has revolutionized exoplanet detection, enabling the discovery of over 5500 confirmed exoplanets via the transit method and around 10000 additional candidates awaiting validation. However, confirming these candidates as true planets demands meticulous vetting and follow-up observations, which hampers the discovery of exoplanets in large-scale datasets. To address this challenge, we developed a machine learning framework trained on Kepler's catalog of confirmed exoplanets and false positives to accurately identify true planetary candidates. Our model uses transit properties, planetary characteristics, and host stellar parameters as training features. The optimized model achieved 83.9\% accuracy in cross-validation. When applied to 3987 TESS candidates with complete observational data, the model identified 2449 high-confidence planets and correctly recovered 86\% (358/418) of all previously confirmed TESS exoplanets in a blinded validation test. Our analysis revealed 100 previously unrecognized multi-planet systems, including five systems that host exoplanets in the habitable zone. Additionally, we identified 15 more planets within the habitable zone of a single system, suggesting strong potential for liquid water stability under conservative planetary albedo assumptions. This work demonstrates that machine learning can accelerate exoplanet validation while maintaining scientific rigor. Our modular design enables direct adaptation to future photometric missions like PLATO or Earth 2.0.
\end{abstract}

\begin{keyword} 
exoplanets \sep machine learning \sep habitability
\end{keyword}

\end{frontmatter}

\section{Introduction}
In the last three decades, significant and rapid advancement has been made in the field of exoplanetary science, starting with the groundbreaking discovery of the first exoplanet orbiting a solar-like star in 1995 through the radial velocity method \citep{1995Natur.378..355M}. A series of space projects with the scientific goal of detecting exoplanets have also emerged, of which NASA's Kepler \citep{2010Sci...327..977B} and Transiting Exoplanet Survey Satellite \citep[TESS,][]{2014SPIE.9143E..20R} space telescopes have contributed their fair share. Both space telescopes use the transit method (as introduced in Section 2.1) to detect the tiny changes in stellar brightness when a planet passes in front of its parent star from an observer's point of view. Additionally, the optical instruments of both telescopes are so advanced that they can simultaneously search for numerous exoplanets with exceptional accuracy.

During its mission, Kepler identified about 2100 eclipsing binaries and 4781 candidate planets, of which 2778 were confirmed as of January 10, 2025. Its 115 square degree field of view covered about 0.25\% of the sky. While Kepler is revolutionary in its finding that Earth to Neptune sized planets were common \citep{2016RPPh...79c6901B}, obtaining ground-based follow-up observations remains challenging due to the majority of stars in the Kepler field being located hundreds to thousands of parsecs away from the solar system. As a result, such follow-up observations are limited to a subset of stars or planet hosts \citep[e.g.,][]{2017AJ....153...71F, 2018ApJ...861..149F}.

To this end, TESS is designed to discover planets orbiting stars within a distance of about 60 parsecs. The objective of the TESS Mission is to survey more than 85\% of the sky, an area 400 times greater than what Kepler covered. Planets detected closer to the Sun are therefore far easier to characterize with follow-up observations, resulting in more refined measurements of planet properties.

Both Kepler and TESS conduct high-precision photometry of the target stars and then identify a list of Threshold Crossing Events (TCEs), which are periodic light flux decrements meeting certain criteria such as a sequence of transit-like features (see Section 2.1). Nevertheless, the task of identifying exoplanets from TCEs is fraught with difficulties since each TCE must undergo thorough scrutiny or "vetting" to eliminate false positives arising from astrophysical variability (such as eclipsing binaries or variable stars), instrumental artifacts, or signal contamination from neighboring stars.

Such vetting was initially conducted manually on an individual basis, which proved to be quite time and resource consuming. Furthermore, the assessment of possible planetary signals by human vetters may lack a consistent adherence to a standardized set of criteria.

Another challenge is the vast volume of data collected through large-scale transit surveys, which requires manual vetting. For instance, Kepler generated hundreds of terabytes of data covering more than 200000 stars, while TESS has the capability to create one million new light curves each month from full-frame images. This has provided us with a never-before-seen amount of photometric observations from space that need to be analyzed efficiently and impartially.

Furthermore, as the number of planet candidates increased, there was a shift in attention towards conducting population-level investigations. This change in focus also resulted in modifications to the methods used for generating and scrutinizing lists of potential planets. Although diverse catalogs that rely on human judgment are effective for prompting subsequent observations, they are unsuitable for population studies, as uniformity is a crucial factor in estimating occurrence rates at the population level. Therefore, there is an urgent need for an efficient, accurate, objective and data-based approach to classify planetary candidates, which is the goal of this work.

To address the limitations of human vetting in transit surveys, several automated tools have been developed to classify light curves. {\scriptsize Robovetter}, developed by \citet{2017ksci.rept....1C}, uses classical tree diagrams and criteria specifically designed to replicate the manual procedure of rejecting false positives. By the end of the Kepler mission, it was fully automated, distinguishing it from machine learning (ML) models, which learn patterns directly from data. \citet{2018AJ....155...94S} marked the first successful application of Convolutional Neural Networks, {\scriptsize AstroNet}, to Kepler data. This model was trained to distinguish exoplanet transits from false positives, achieving a high accuracy of 98.8\% in ranking planetary signals above false positives. Building on {\scriptsize AstroNet}, \citet{2019AJ....158...25Y} developed {\scriptsize AstroNet-Triage} and {\scriptsize AstroNet-Vetting}, the first models to use real TESS data for training and testing. {\scriptsize AstroNet-Triage} focuses on filtering out obvious false positives while retaining planetary candidates, achieving a 93\% reduction in non-planetary signals without compromising completeness. {\scriptsize AstroNet-Vetting} further aims to differentiate eclipsing binaries from planetary candidates but has shown lower precision in its early stages, with ongoing efforts to refine its performance. \citet{2022NewA...9101693O} applied a novel ML system, developed by ThetaRay, Inc.\footnote{\url{https://thetaray.com}}, trained on Kepler data and validated with confirmed exoplanets, to analyze 10,803 TESS TCEs. Using semi-supervised and unsupervised techniques, they identified 50 high-probability candidates and uncovered three new exoplanets, demonstrating the first successful application of combined ML methodologies for rapid classification in large astrophysical datasets. \citet{sturrock2019} applied ML for exoplanet classification, training several classification models with transit properties obtained from the Cumulative Kepler object of interest (KOI) table\footnote{\url{https://exoplanetarchive.ipac.caltech.edu/cgi-bin/TblView/nph-tblView?app=ExoTbls&config=cumulative}}. Their best-performing model, a Random Forest classifier, achieved a 98\% cross-validated accuracy and identified 968 high-probability candidates ($>95{\%}$). However, their study was limited to the Kepler dataset, as many of the training features are unavailable in TESS, and it did not incorporate stellar parameters. \citet{2024AJ....168..100T} developed a one-dimensional convolutional neural network (CNN) to differentiate real transit events from false positives in TESS light curves, reducing the need for manual vetting while maintaining high accuracy in planet candidate classification.

In this work, we aim to develop an ML model that can efficiently and accurately identify exoplanet candidates using TCE features as well as stellar parameters that are relatively easy to obtain from observations. The model will be trained on the available data of confirmed exoplanets from Kepler and then applied to the TESS candidates to identify new exoplanet candidates. The paper is organized as follows: Section 2 provides introductions of the transit method, the ML models and the evaluation metrics used in this work; Section 3 details the data preparation, ML model training, and testing and evaluation; Section 4 presents the use of the trained model to classify TESS exoplanet candidates; Section 5 discusses the results and implications of this work; And Section 6 concludes the paper.

\section{Background and Methodology}
\subsection{Exoplanet Transits}

The data used in this project came from the Kepler and TESS space telescopes, both of which observe small changes in the brightness of host stars during transits. The brightness of the host star observed  will temporarily decrease when a planet passes between the star and the observer. This phenomenon, known as a planet transit. When the planet moves out of the star's disk, the brightness returns to normal.

During the transit, the change in the star's brightness (flux), defined as $\Delta F$, is directly related to the size of the planet. This relationship can be expressed using the following formula: 

\begin{equation}
\frac{\Delta F}{F} = \frac{R_\mathrm{p}^2}{R_*^2}
\label{eq:transit_depth}
\end{equation}

Where $F$ is the total brightness (flux) of the host star when it is not occulted, and the $R_p$ and $R_*$ are the radius of the planet and the host star, respectively. Different sized planets will block different amounts of light as they transit, which is reflected in the depth of the dip in the light curve, i.e. transit depth expressed as $\frac{\Delta F}{F}$. Therefore, the size ratio of the planet to its host star can be calculated by measuring the change in brightness of the star during the transit (Equation 1). Then, if the radius of the host star is obtained through other means, the radius of the planet can be further determined. Another important parameter is the transit duration  $T_{dur}$, which measures the length of the planet's transit, related to parameters such as the planet size, period, and orbit. These parameters are important criteria to distinguish planets from stars. For instance, the size of a planet should be significantly smaller than a star. 

The advantage of the transit method is that the transits can be observed repeatedly and therefore can be checked and analyzed repeatedly. In addition, telescopes can monitor the transit events of many stars simultaneously and is currently the most commonly used method to find exoplanets. On the other hand, the velocity method usually relies on spectrum observations on the ground and therefore it is affected by sunlight, weather, light pollution, etc . and it is usually not possible to observe radial velocity for many stars at the same time. But spectroscopic observations can still help astronomers to detect and confirm exoplanets \citep{2016ASSL..428....3H}, to determine planet masses and orbital parameters \citep{2021AJ....162..266L}, and to observe the composition of planet atmospheres \citep{2024NatSR..1427356R}. 

\subsection{Methodology}
\subsubsection{Model Training Design}
Therefore, how to find the real planet from a large amount of transit observation data is an urgent problem to be solved. As artificial intelligence  is utilized across diverse domains of scientific research, ML can also be employed to assist in the discovery of exoplanets. In this work, we aim to use transit data as training samples for ML, and the trained model is then used to identify new exoplanet candidates. For comparison, we will examine the results of 5 different algorithms that take the same training samples from Kepler and finally select the best model to predict TESS dataset. 

The 5 different algorithms include a Transformer-based neural network as well as four machine learning algorithms: logistic regression, K-nearest neighbors, decision tree and random forest. All models are trained on a common dataset to ensure consistent evaluation and comparison.

The training set is obtained from the Kepler mission and includes the properties of exoplanets and their host stars. This dataset offers a large number of exoplanet candidates with precisely measured TCE properties, including both confirmed candidates and labeled false positives. The models are trained on numerical features of the TCEs rather than light curves, as our focus is on analyzing structured data rather than performing image processing of the light curves or time-series analysis. A detailed description of the dataset and preparation steps are given in Sections~\ref{sc:data} and~\ref{sc:data_prep}.

After training, each model is evaluated, compared, and fine-tuned through hyperparameter optimization. The best-performing model is then used to identify exoplanet candidates in the TESS dataset, where less than 10\% of the current (TESS object of interest) TOI planets have been validated. The selected candidates are further analyzed based on the distributions of their planetary and stellar parameters. A prioritized list of high-interest candidates is generated to guide future follow-up observations. Additionally, the habitability potential of these exoplanets is assessed by determining whether they reside within the habitable zone of their host stars.

\subsubsection{Evaluation Metrics}
Each of the ML models will be evaluated on the following four metrics:

\begin{equation}
    \begin{aligned}
        &\mathrm{Accuracy} = \frac{\mathrm{TP} + \mathrm{TN}}{n}, \\
        &\mathrm{Precision} = \frac{\mathrm{TP}}{\mathrm{TP} + \mathrm{FP}}, \\
        &\mathrm{Recall} = \frac{\mathrm{TP}}{\mathrm{TP}+ \mathrm{FN}}, \\
        &\mathrm{F1\,Score} = 2*\frac{\mathrm{Precision} * \mathrm{Recall}}{\mathrm{Precision} + \mathrm{Recall}},
    \end{aligned}
    \label{eq:metrics}
\end{equation}
where $n$ is the number of predicted samples, TP is the number of true positives, TN is the number of true negatives, FP is the number of false positives, and FN is the number of false negatives, which are the four possible outcomes of a binary classification problem. The ideal outcome for ML models is to minimize both false negatives and false positives.

The implications of the 4 evaluation metrics are as follows:
\begin{itemize}
    \item Accuracy is the measurement of correctly predicted outcomes relative to all predictions made.
    \item Precision refers to the proportion of accurately predicted positives by each model out of all the positives that were predicted. 
    \item Recall can be defined as the ratio of correctly identified positive instances to all actual positive instances.
    \item F1 Score is calculated as a harmonic mean of precision and recall. 
\end{itemize}

In this work, the best-performing model will be selected based on its evaluation and used to predict planets for TOI targets.

\subsection{Data}
\label{sc:data}
We adopt global parameters of the host stars of the exoplanet candidate as well as transit properties of all TCEs collected in the Kepler Mission as the training set, analyze and compare the performance of multiple ML models under the same training data, and finally selects the optimal model, and then applies it to the TOI table to identify signals from exoplanets. Despite the different observation strategies and satellite payloads, the TESS Mission greatly resembles Kepler in terms of detection method, data quality, data processing techniques, etc. Thus, the model trained with KOI properties should also work to identify planet candidates in the TOI table.

As of January 10, 2025, Kepler found 9,564 KOIs that were suspected of hosting one or more transiting planets. KOIs come from a master list of 150,000 stars, which itself is generated from the Kepler Input Catalog (KIC). A KOI shows a periodic dimming, indicative of an unseen planet passing between the star and Earth, eclipsing part of the star. However, such an observed dimming is not a guarantee of a transiting planet. Due to this reason, most KOIs are still unconfirmed as transiting planetary systems.

As for TESS, there are currently a total of 7358 TOI targets. Excluding those with missing data in any of the selected features, the dataset has a total of 6330 TOIs available for further ML prediction. TOIs are the first step in the confirmation process of exoplanet candidates, and the majority of them have not been confirmed as planets or false positives. The best-performing model will be used to predict the TOI data and identify new exoplanet candidates.

\begin{table}[h!]
    \centering
    \caption{Stellar and planetary parameters used to train the ML models.} \label{tb:data_columns}
    \renewcommand{\arraystretch}{1.5}
    \begin{tabularx}{\textwidth}{|>{\centering\arraybackslash}m{0.25\textwidth}|>{\RaggedRight}X|}
    \hline
    \multicolumn{1}{|c|}{\textbf{Features}} & \multicolumn{1}{c|}{\textbf{Description}} \\ \hline
    $P$ (d) & The orbital period of the planet. \\ \hline
    $t_0$ (BKJD) & The time corresponding to the center of the first detected transit. \\ \hline
    $t_\mathrm{dur}$ (h) & The duration of the observed transits.\\ \hline
    $\delta$ (ppm) & Transit depths. \\ \hline
    $R_\mathrm{p}$ ($\mathrm{R}_\oplus$) & The radius of the planet. \\ 
    \hline
    $T_\mathrm{p}$ (K) & The temperature of the planet. \\ \hline
    $S$ ($\mathrm{S}_\oplus$) & Insolation flux, in unit of earth insolation. \\ \hline
    $T_\mathrm{eff}$ (K) & The photospheric temperature of the star. \\ 
    \hline
    $\log g$ & The stellar surface gravity, $g$ is in $\mathrm{cm}/\mathrm{s}^{2}$. \\ 
    \hline
    $R_\star$ ($\mathrm{R}_\odot$) & The photospheric radius of the star. \\ \hline
    [Fe/H] & The stellar metallicity. \\ \hline
    \end{tabularx}
\end{table}

From the KOI and TOI tables, several transit-related parameters are obtained. In addition, a number of host star features will  also be used. These stellar parameters are effective temperature, surface gravity, radius, and metallicity.
The stellar and planetary parameters used to train the ML models are listed in Table~\ref{tb:data_columns}\footnote{Detailed definitions of the variables listed in Table 1 can be found in the NASA Exoplanet Archive: \url{https://exoplanetarchive.ipac.caltech.edu/docs/data.html}}. 

Stellar parameters, particularly metallicity, play a crucial role in improving the classification of planetary candidates and understanding planet formation. Studies such as \citet{2012Natur.486..375B, 2015AJ....149...14W, 2014Natur.509..593B} have demonstrated a strong correlation between host star metallicity and planet occurrence rates. Higher metallicity is associated with increased planet formation, especially for gas giants, aligning with the core accretion theory of planetary formation. This theory posits that metallicity influences the growth of planetary cores and their ability to accrete gas before its dispersal, particularly in environments with varying metallicity. However, it is possible that this trend of planets being more common around stars with higher metallicity may be partly influenced by observational bias. Compared to our Milky Way galaxy, 9564 stars is a very small sample size, and the observation targets of Kepler and TESS are in the close vicinity of the Sun, with a population of relatively young stars with relatively high metallicity. But given its high correlation to planet occurrences, it is still worth including metallicity as an additional training feature in the training. 

\section{Model Training, Testing and Evaluation}
\subsection{Kepler Data Preparation} \label{sc:data_prep}
The training dataset, the cumulative KOI data table, is downloaded from NASA's Exoplanet Online Archive\footnote{\url{https://exoplanetarchive.ipac.caltech.edu}}. 
The complete KOI catalog contains more than 80 feature columns, from which we selected the 11 columns listed in Table~\ref{tb:data_columns} as training features.
These features comprise of 7 transit-related and 4 stellar parameters. 

\begin{figure}[h!]
    \centering
    \includegraphics[width=\textwidth]{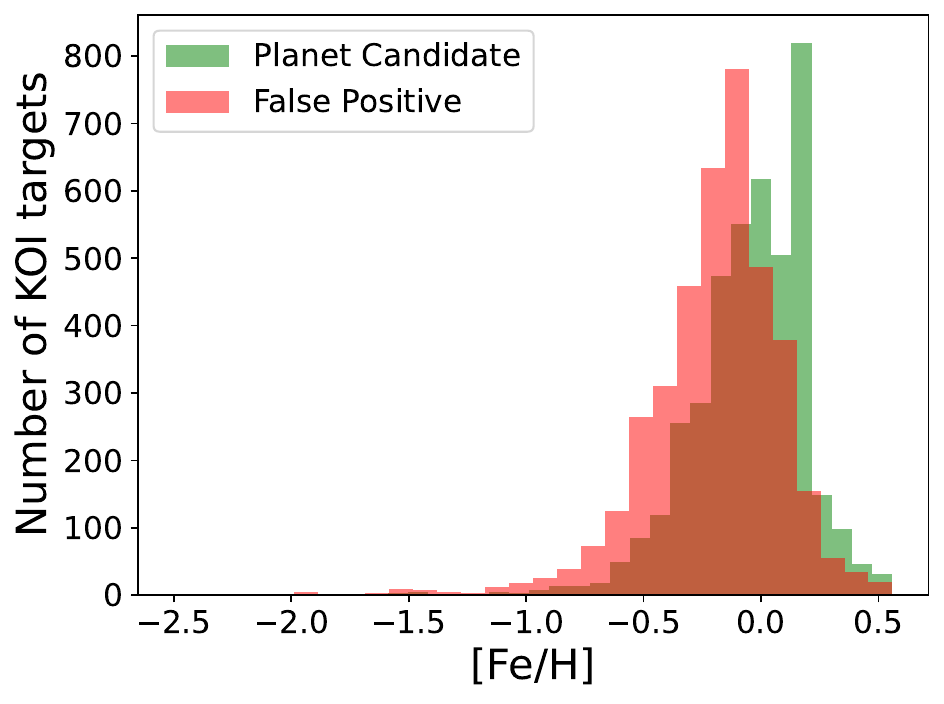}
    \caption{Histogram of Metallicity from KOI Table.}
    \label{fg:koi_metallicity}
\end{figure}

A brief population analysis on the KOI dataset shows that a majority of planet candidates have higher metallicities, while a majority of false positives have lower metallicity (Figure~\ref{fg:koi_metallicity}). Such a distribution pattern clearly supports the correlation between metallicity of the host star and occurence of planets.

The \texttt{koi\_pdisposition} column in the KOI table provides the physical classification for each KOI. Specifically, entries labeled CANDIDATE in this column are assigned 1 (planet candidates), while those labeled FALSE POSITIVE are assigned 0 (false positives).

\begin{figure}[h!]
    \centering
    \includegraphics[width=\textwidth]{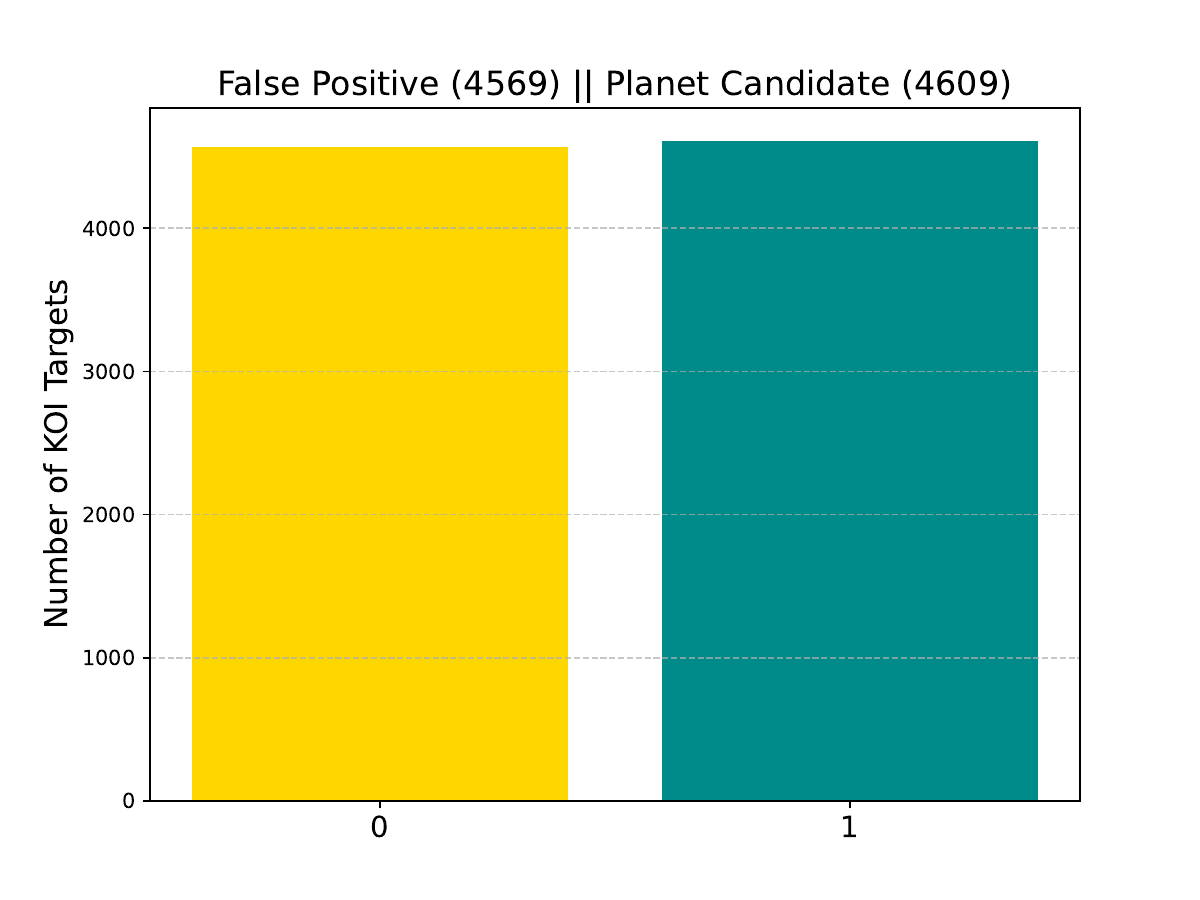}
    \caption{Number of KOI targers that are classified as false positive and planet candidates in the training sample.}
    \label{fg:koi_data_ratio}
\end{figure}

After dropping the targets with NaN value in any of the feature and label columns, 9178 KOI targets are available for training the ML models, out of which the ratio between planet candidates and non-planets is close to 1:1 (see Figure~\ref{fg:koi_data_ratio}). This shows there is no obvious bias on the data labels, which is ideal as a training sample for ML.

The KOI data is divided into a training set (80\%) and a testing set (20\%) randomly. The first one is employed for training each ML model whereas the second one is employed for evaluating the performances. The process of data distribution is a pseudo-random operation, that is, the same random seed is being used for each allocation to guarantee that the data distribution itself is random and the data used by different ML algorithms is the same.

\subsection{Training models}
\label{sc:train_model}
The KOI training datasets prepared in Section~\ref{sc:data_prep} are fed into four independent ML algorithms, including Logistic Regression, K-Nearest Neighbors Classification, Decision Tree, and Random Forest, as well as a transformer encoder-based neural network. Each of the above models is appropriate for classification problems with its own unique advantages. In this work, we will also compare the performance of each model based on the training and test results of KOI data.

The first four ML algorithms are implemented using the open-source Python library {\scriptsize scikit-learn}\footnote{\url{https://scikit-learn.org/}} and executed on Jupyter Notebook\footnote{\url{https://jupyter.org}}. The data preprocessing uses StandardScaler in {\scriptsize scikit-learn} to normalize the data to ensure that each algorithm can converge stably. Each model is trained on a single CPU on a local computer, and depending on the type of model, the training time varies from 30 seconds to 5 minutes. 

In order to find the most optimal hyperparameters for each model, we have conducted a series of experiments and trials. For each model, the hyperparameter with the greatest impact on the prediction result is selected and then the prediction results are analyzed by changing the input of the selected hyperparameters. The hyperparameters with the highest accuracy in prediction results will be chosen as the optimal input for the particular model. In Table~\ref{tb:all_hyperparameters}, the hyperparameters for each model are listed, together with the key parameter that has the greatest impact on the prediction result. The range of the key hyperparameters that are tested is also provided.

\begin{table}[h!]
    \centering
    \caption{Hyperparameters adopted for the ML models. In the last column are the ranges of the key hyperparameters that have been tested.}
    \label{tb:all_hyperparameters}
    \begin{tabular}{|l|c|c|}
        \hline
\        Model & Hyperparameters & Key parameter Range \\
        \hline
        Decision Tree & \makecell[c]{\textit{max\_depth}=5\\ \textit{min\_samples\_split}=3 \\ \textit{min\_samples\_leaf}=6} & \textit{max\_depth}: 1--10 \\
        \hline
        K Nearest Neighbor & \makecell[c]{\textit{n\_neighbors}=12\\ \textit{leaf\_size}=30\\ \textit{metric}=`Manhattan'\\ \textit{weights}=`uniform'} & \textit{n\_neighbors}: 10--30 \\
        \hline
        Random Forest & \makecell[c]{\textit{n\_estimators}=340\\ \textit{criterion}=`gini'\\ \textit{bootstrap}=False\\ \textit{min\_samples\_split}=10\\ \textit{min\_samples\_leaf}=8} & \textit{n\_estimators}: 100--500 \\
        \hline
        Logistic Regression & \makecell[c]{\textit{C}=100\\ \textit{max\_iter}=100\\ \textit{tol}=0.01\\ \textit{class\_weight}=`balanced'} & \textit{C}: 100--200 \\
        \hline
    \end{tabular}
\end{table}

\begin{figure}[h!]
    \centering
    \includegraphics[width=\textwidth]{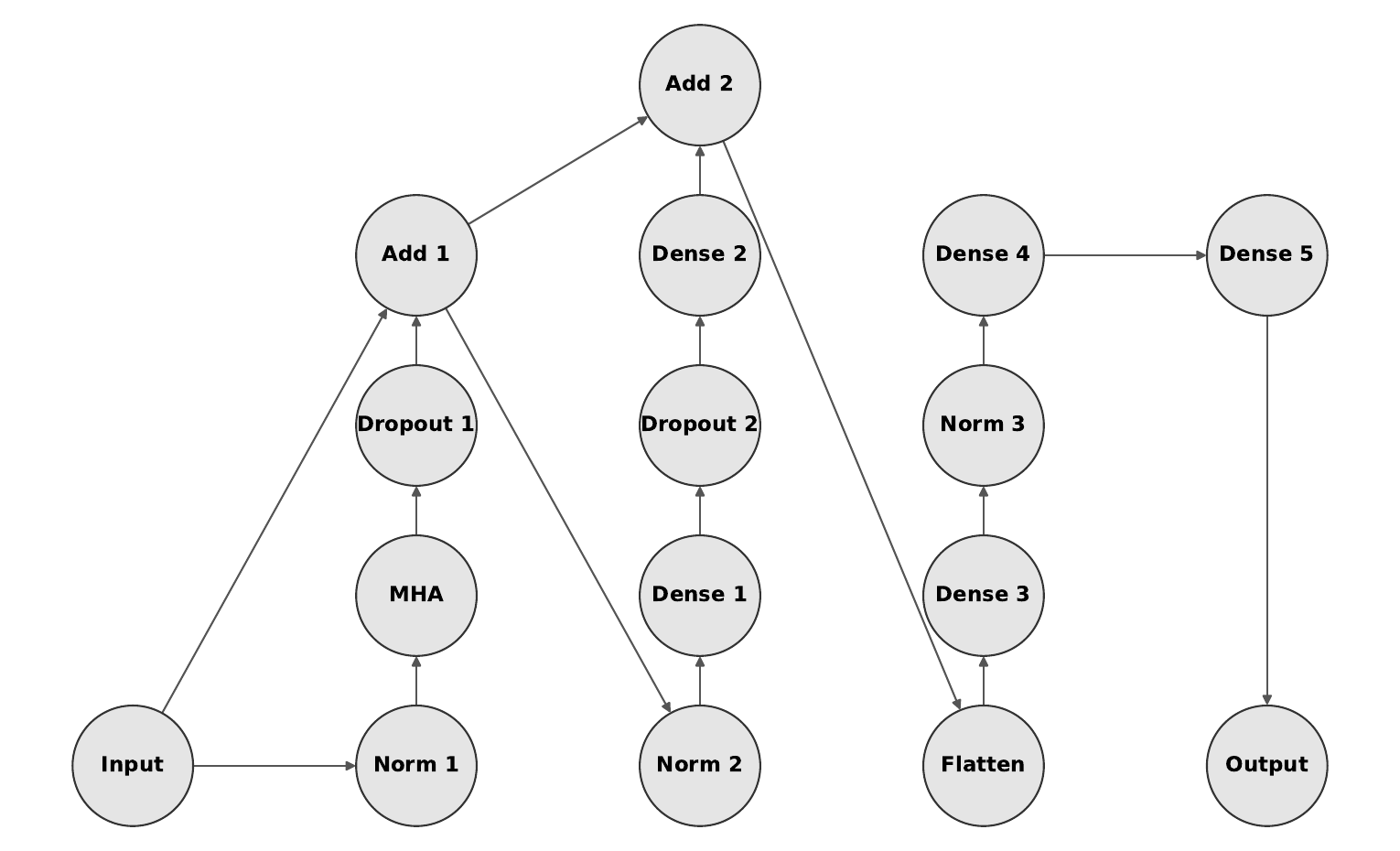}
    \caption{Architechture of the neural network model. Definition and shape of each layer are provided in Table~\ref{tb:neural_network_layers}.}
    \label{fg:neural_network}
\end{figure}

Slightly different from the previous four ML models, the transformer encoder\textendash{}based neural network is a type of deep learning model.
The transformer encoder is a specific neural network architecture designed to process sequential data by capturing contextual relationships within the input. 
It relies on self-attention mechanisms to weigh the importance of different elements in the sequence, enabling it to model complex dependencies and patterns. The encoder consists of multiple layers, each containing multi-head self-attention and feed-forward neural networks, followed by layer normalization and residual connections. 
These components work together to transform the input features into a rich, context-aware representation. For a binary classification task to predict whether a KOI is a true planet or a false positive, the transformer encoder will be applied to the 11 selected training features, The encoder processes these features as a sequence, leveraging self-attention to identify relationships and patterns among them. The output representation is then passed through a classification head, a fully connected layer with a sigmoid activation, to produce a probability score for the binary outcome. By training the model on labeled KOI data, the transformer encoder learns to distinguish between true planets and false positives, making it a powerful tool for exoplanet discovery and validation. Table~\ref{tb:neural_network_layers} provides the details of the neural network layers, while Figure~\ref{fg:neural_network} illustrates the architecture and connections between the layers.
\begin{table}[h!]
    \centering
    \caption{Layers of the neural network.} \label{tb:neural_network_layers}
    \begin{tabular}{@{}lcc@{}}
        \toprule
        \textbf{Layer (type)} & \textbf{Output Shape} & \textbf{Param \#} \\ 
        \midrule
        Input Layer (Input) & (None, 11, 1)  & 0    \\
        Layer Normalization (Norm 1) & (None, 11, 1)  & 2   \\
        Multi--Head Attention (MHA) & (None, 11, 1)  & 449  \\
        Dropout 1 & (None, 11, 1)  & 0    \\
        Add 1 & (None, 11, 1)  & 0   \\
        Layer Normalization (Norm 2) & (None, 11, 1)  & 2   \\
        Dense 1 & (None, 11, 64) & 128   \\
        Dropout 2 & (None, 11, 64) & 0   \\
        Dense 2  & (None, 11, 1)  & 65 \\
        Add 2  & (None, 11, 1)  & 0   \\
        Flatten   & (None, 11)     & 0     \\
        Dense 3 & (None, 32)   & 384    \\
        Layer Normalization (Norm 3) & (None, 32)     & 64   \\
        Dense 4 & (None, 16) & 528   \\
        Dense 5  & (None, 8)  & 136 \\
        Output Layer (Output)  & (None, 1)      & 9     \\ 
        \midrule
        \multicolumn{3}{@{}l}{\textbf{Total params:} 1,767 (6.90 KB)} \\
        \multicolumn{3}{@{}l}{\textbf{Trainable params:} 1,767 (6.90 KB)} \\
        \multicolumn{3}{@{}l}{\textbf{Non-trainable params:} 0 (0.00 B)} \\
        \bottomrule
    \end{tabular}
\end{table}

To train the transformer, we adopted a learning rate of 0.001, a batch size of 8, and the Adam optimizer. In addition, we tested the model with epochs of 100 to 400 and the final number of epochs is set at 200 as the model delivered the highest training accuracy. The model is trained on a single GPU on a local computer, and the training time is approximately 17 minutes.

All five models are evaluated on the KOI test set to assess their performance in predicting exoplanet candidates. The results are presented in the following section.

\subsection{Testing and Evaluation}

\begin{figure}
    \centering
    \includegraphics[width=\textwidth]{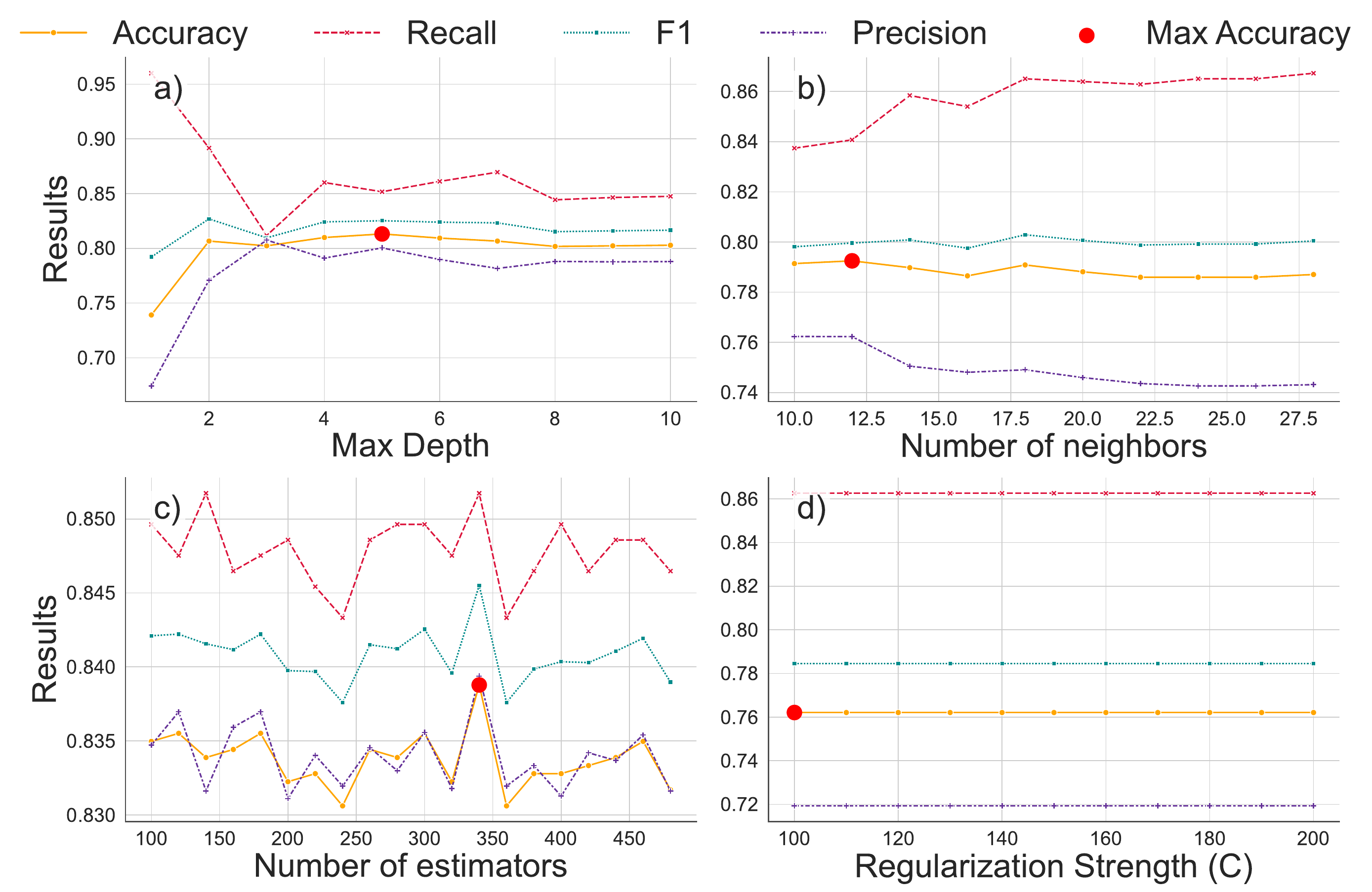}
    \caption{Impact of hyperparameters on the performance of the four models: a) Decision Tree, b) K Nearest Neighbors, c) Random Forest,and d) Logistic Regression. In each subfigure, the most optimal hyperparameter is marked with a red dot.}
    \label{fg:hyperparameters}
\end{figure}

The impact of key hyperparameters on the prediction results for each model is evaluated and compared, using the metrics defined in Equation~\ref{eq:metrics}. Figure~\ref{fg:hyperparameters} shows the comparisons of the prediction results for each model with different hyperparameters.

For Decision Tree, the model has the highest accuracy when \textit{max\_depth} is 5 which characterizes the number of nodes of the decision tree used. For the K Nearest Neighbor Classification, the \textit{number\_of\_neighbors} is particularly important because it mainly relies on the "neighbor's voting mechanism". When \textit{number\_of\_neighbors} = 12, the model performs the best. For Random Forest, the monitoring hyperparameter \textit{n\_estimatiors} is selected, which is roughly equivalent to the number of decision trees utilized by the model. According to the evaluations, the Random Forest achieves its highest accuracy when the number of estimators is set to 340. Finally, for Logistic Regression, as shown in panel d of Figure~\ref{fg:hyperparameters}, we did not gain any substantial benefit while tuning this hyperparameter, and the classifier with default values (100) of $C$ was strong enough by itself.

The test data set, which comprises 20\% of the KOI data not utilized for training, is employed to analyze and evaluate the performance of the 5 algorithms already trained. The predictions of each model are then checked against the labels of the corresponding test set KOI. The evaluation is based on the accuracy of each model's predictions of planet candidates as well as its performance in terms of efficiency and stability. The 4 metrics (Equations 2) are used to conduct such assessments. The specific results are presented in Table~\ref{tb:performance}. It should be noted that the results in the table are achieved after optimizing the input hyperparameters, as indicated in the middle column of Table~\ref{tb:all_hyperparameters}.

\begin{table}[h!]
    \centering
    \caption{Performance Metrics and Hyperparameters for Various Algorithms}
    \label{tb:performance}
    \resizebox{\textwidth}{!}{
    \begin{tabular}{l c c c c c c c c c}
    \toprule
    Algorithm & Precision & Recall & Accuracy & F1 & TN & FP & FN & TP \\
    \midrule
    Decision Tree & 0.800 & 0.852 & 0.813 & 0.825 & 683 & 202 & 141 & 810 \\
    K Nearest Neighbor & 0.762 & 0.841 & 0.792 & 0.800 & 695 & 237 & 144 & 760 \\
    Random Forest & 0.839 & 0.852 & 0.839 & 0.846 & 730 & 155 & 141 & 810 \\
    Logistic Regression & 0.719 & 0.863 & 0.762 & 0.784 & 606 & 311 & 127 & 797 \\
    Transformer & 0.815 & 0.860 & 0.826 & 0.837 & 699 & 186 & 133 & 818 \\
    \bottomrule
    \end{tabular}
    }
\end{table}

Based on the metrics in Table~\ref{tb:performance}, the following conclusions can be drawn:

\begin{itemize}
    \item The Random Forest model exhibits the most robust performance among all the algorithms evaluated. It boasts high precision (0.839), recall (0.852), and accuracy (0.839), alongside an impressive F1 score of 0.846. These figures indicate its exceptional ability to identify true positives while maintaining a high level of accuracy in its predictions. 
     \item The Transformer neural network model follows closely behind Random Forest, achieving the second-highest overall accuracy (0.826) and the F1 score (0.837). Despite its current slight underperformance relative to the Random Forest, the Transformer's potential for improvement is significant. Given the current training set's limited size of around 7000 targets, the neural network's performance is understandably somewhat constrained. However, as the dataset expands to include a greater variety of exoplanets with diverse characteristics, the Transformer is poised to enhance its capabilities and potentially surpass the Random Forest in future applications.   
    \item The Decision Tree model demonstrates commendable performance, with all key metrics exceeding 80\%. This indicates a solid overall effectiveness, though it does not quite match the superior results of the Random Forest and Transformer. 
    \item The K Nearest Neighbor Classification and Logistic Regression models exhibit comparatively weaker overall performance. Their lower precision, recall, accuracy, and F1 scores suggest that these algorithms struggle with datasets containing a large number of less distinct features. Consequently, they are not the most suitable choices for this work.
\end{itemize}

Based on the analysis, the Random Forest model is selected to predict planets for TOI targets because of its superior performance across key metrics. The Transformer model will serve as a backup if more diverse training data becomes available in the future. The Decision Tree, despite decent performance, is not used as it is generally similar to the Random Forest. The K Nearest Neighbor and Logistic Regression models are excluded due to their poor performance.

\begin{figure}
    \centering
    \includegraphics[width=\textwidth]{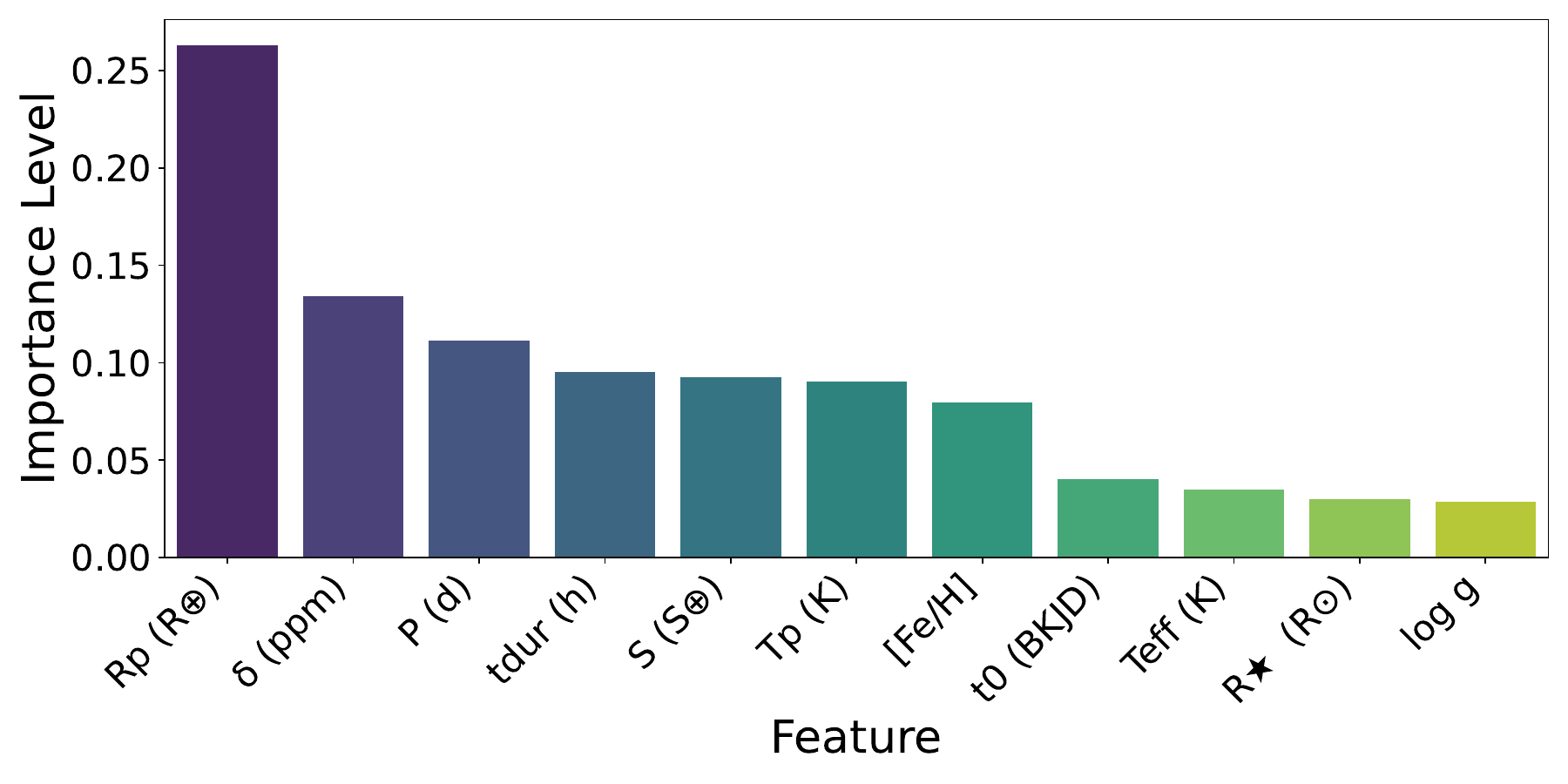}
    \caption{Feature importance chart for the Random Forest model.}
    \label{fg:feature_importance}
\end{figure}

The importance of the selected training features for planet classification can be assessed using the Random Forest model. As illustrated in Figure~\ref{fg:feature_importance}, the planet radius ($R_\mathrm{p}$) emerges as the most crucial feature, followed by transit depth $\delta$, orbital period ($P$), and transit duration ($t_\mathrm{dur}$). These four features exhibit similar levels of importance.

It is noteworthy that, theoretically, the mass of a celestial body is the primary determinant of whether it qualifies as a planet. However, planet mass cannot be directly measured by instruments like Kepler or TESS and requires supplementary methods such as radial velocity observations. Consequently, only a limited number of KOI objects, approximately 300, have mass data available. Therefore, mass is not included as a training feature in this work. Despite the absence of mass data, the radius of an object remains a vital parameter, especially for smaller celestial bodies, which are more likely to be planets. The prominence of radius in the feature importance ranking supports its significance in distinguishing planetary objects. This finding aligns with the broader understanding that both mass and radius are essential for accurate planetary classification, with radius serving as a practical substitute in the absence of mass measurements.

Furthermore, although metallicity ([Fe/H]) does not rank among the top three features, it still holds a significant position with an importance level close to 0.10. Notably, it is the most important stellar parameter among the four included in the training dataset. This aligns with the established strong correlation between stellar metallicity and planet occurrence, indicating that metallicity plays a crucial role in planet formation processes.

\section{Identifying TESS Exoplanet Candidates}
\label{sc:toi}
\subsection{TESS Data Preparation}
\label{sc:tess_data_prep}
The TOI dataset were extracted following the same procedure outlined in Section~\ref{sc:data_prep} for the KOI dataset. As of January 10, 2025, the NASA Exoplanet Online Archive comprises a total of 7358 TOI targets. After excluding those with missing data in any of the selected features, the refined dataset comprises 6330 TOIs, which are available for further machine learning prediction.

Unlike the KOI table, [Fe/H] is unavailable in the TOI table. In this study, we obtained [Fe/H] for TOI targets from Gaia Data Release 3 \citep[DR3,][]{2023A&A...674A...1G}. We first performed a cross-match between the TOI table and the Gaia DR3 table by matching the identifiers of the targets in SIMBAD\footnote{\url{https://simbad.cds.unistra.fr/simbad/}}. Through this process, we successfully identified Gaia source IDs for 4606 out of the 6330 TOI targets.

Subsequently, we searched for metallicity in the Gaia DR3 table from the Gaia Archive\footnote{\url{https://gea.esac.esa.int/archive/}}. Since [Fe/H] is not directly avaliable in Gaia DR3, we have to retrieve the total metallicity ([M/H]) instead. [M/H] was available for 4259 TOIs, derived either from spectroscopy \citep[GSP-Spec,][Gaia data column \texttt{mh\_gspspec}]{2023A&A...674A..29R} or, if unavailable, from photometry  \citep[GSP-Phot,][Gaia data column \texttt{mh\_gspphot}]{2023A&A...674A..27A}.

To convert [M/H] to [Fe/H], we applied three different methods sequentially:
{\begin{enumerate}
    \item \textbf{Iron-to-metal ratio conversion}: We used the abundance of neutral iron ([Fe/M]) from GSP-Spec (Gaia data column \texttt{fem\_gspspec}) and calculated [Fe/H] as:
    \[
    \text{[Fe/H]} = \text{[Fe/M]} + \text{[M/H]}.
    \]
    This method yielded [Fe/H] for 1339 TOIs.
    \item \textbf{Alpha-element correction}: If the first method was not possible, we used the abundance of alpha elements ([$\alpha$/Fe], Gaia data column \texttt{alphafe\_gspspec}) with the empirical relation:
    \[
    [\text{Fe/H}] = [\text{M/H}] - \log_{10} \left( 0.694 \times 10^{[\alpha/\text{Fe}]} + 0.306 \right).
    \]
    This method provided metallicities for an additional 1320 TOIs\footnote{We note that this relation is tailored for stars with significant alpha-element enrichment.}.
    \item \textbf{Photometric calibration}: If neither spectroscopic method was applicable, we used the calibrated GSP-Phot metallicities via the Python tool {\scriptsize gdr3apcal}\footnote{\url{https://github.com/mpi-astronomy/gdr3apcal}}. This approach yielded [Fe/H] for 1328 TOIs.

\end{enumerate}
In total, we obtained [Fe/H] data for 3987 TOI targets, which were used for the subsequent ML prediction.


\subsection{TOI Prediction and Evaluation}
\label{sc:toi_prediction}
The Random Forest model, trained in Section~\ref{sc:train_model}, was applied to the 3987 TOI targets for which all training features were readily available. The model successfully identified 2449 potential planet candidates. These candidates were then cross-referenced with the TESS Follow-up Observing Program Working Group (TFOPWG) disposition results. Specifically, the TFOPWG has categorized 977 of these TOI targets as either CP (Confirmed Planet) or FP (False Positive), as delineated in the \texttt{TFOPWG\_Disposition} column. To assess the model's performance, a blinded validation test was conducted using the \texttt{TFOPWG\_Disposition} labels for these 977 TOIs. The results were as follows:
\begin{itemize}
    \item A total of 418 TOI targets were labeled as CP, of which 361 were accurately identified as planet candidates by our model, thereby achieving an impressive accuracy of 86.4\%.
    \item Conversely, 559 TOI targets were labeled as FP, of which 384 were successfully excluded by our model, resulting in an accuracy of 68.7\%.
\end{itemize}
When considering both CP and FP categories together, the model demonstrated a recall of 86.4\% and an overall accuracy of 76.3\%, representing a significant 11\% improvement over the {\scriptsize AstroNet-Vetting} \citep{2019AJ....158...25Y} which also utilized TESS data. For a comprehensive overview, Table~\ref{tb:toi_prediction} provides the detailed prediction outcomes for all 3987 TOI targets, including the number of NP (Non-planet) and Planet predictions, as well as the accuracy of the model against the 977 TOI targets that have been classified as CP and FP by the TFOPWG.

\begin{table}[ht]
    \caption{TOI Prediction Results. The table shows the classification of TOI targets based on TFOPWG dispositions (TOI column). The Random Forest and Transformer models were used to predict whether the TOI targets were planet candidates (Planet column) or non-planets (NP column). The predictions from both models were then categorized according to the TFOPWG dispositions for comparative analysis. 
    Model accuracy was calculated for CP and FP categories.
    }
    \label{tb:toi_prediction}
    \centering 
    \resizebox{\textwidth}{!}{
    \begin{threeparttable}
        \begin{tabular}{ccccccccc} %
        \toprule
        \multirow{2}{*}{Dispositions}
        & \multirow{2}{*}{TOI} &
        \multicolumn{3}{c}{Random Forest} & &
        \multicolumn{3}{c}{Transformer}\\ 
        \cmidrule{3-5} \cmidrule{7-9}
        &  &{NP} & {Planet} & Accuracy 
        & & {NP} & {Planet} & Accuracy\\
        \midrule
        APC & \small{207}  & \small{151} & \small{56} & & & \small{117} & \small{90} & \\
        CP  & \small{418} & \small{57} & \small{361} & \small{86.4\%}  & & \small{209} & \small{209} & \small{50.0\%} \\
        FA  & \small{45} & \small{17} & \small{28} &&&  \small{26} & \small{19} & \\
        FP  & \small{559} & \small{384} & \small{175} & \small{68.7\%} & &  \small{368} & \small{191} & \small{65.8\%} \\
        KP  & \small{501} & \small{211} & \small{290} & & & \small{295} & \small{206} & \\
        PC  & \small{2257} & \small{718} & \small{1539} & & & \small{1010} & \small{1247} & \\
        \midrule
        Total & \small{3987} & \small{1538} & \small{2449} &&&  \small{2025} & \small{1962} &\\
        \bottomrule
        \end{tabular}
        \begin{tablenotes}
            \setlength\labelsep{0pt}
            \scriptsize
            \item \textbf{Notes.} {APC: ambiguous planetary candidate, CP: confirmed planet. FA: false alarm, FP: false positive, KP: known planet, PC: planetary candidate, NP:Non-planet.}
        \end{tablenotes}
    \end{threeparttable}
    }
\end{table}

The Transformer model was also employed to classify the same set of TOI targets, and its prediction results for the 977 TOI targets were compared against the TFOPWG disposition results (see also Table~\ref{tb:toi_prediction}). The performance of the Transformer model on TOI targets labeled as FP was comparable to that of the Random Forest model. However, its performance on TOI targets labeled as CP was notably inferior.

A detailed analysis of the Transformer model predictions for the 418 CP TOI targets revealed significant disparities. Specifically, only 45.6\% of the CP TOI targets with $P$ less than 14 days were correctly identified. In contrast, the prediction accuracy improved to 64.3\% for CP TOI targets with orbital periods of more than 14 days.

This discrepancy can be largely attributed to the significant differences in the distribution of orbital periods between the KOI training datasets and the TOI targets used for prediction. Only 58.9\% of the KOI targets in the training datasets have a $P$ less than 14 days, whereas this percentage increased to 84.7\% for the TOI targets available for prediction. This mismatch in the distribution of $P$ between the training and prediction datasets likely contributed to the Transformer model's lower performance on shorter-period CP TOI targets. For a more in-depth analysis of these differences between TOI and KOI datasets, refer to Section~\ref{sc:tess_population}.

In addition, to further evaluate the impact of [Fe/H] on the prediction results, we retrained the Random Forest model using the same set of training features, but this time excluding [Fe/H]. The model was then applied to the 3987 TOI targets, and the results were compared with those obtained from the model that included [Fe/H] as a training feature. 
Figure~\ref{fg:metallicity_histogram} shows the side-by-side comparison of the distribution of [Fe/H] for the TOI targets classified by the Random Forest model with and without [Fe/H]. The results indicate that the model trained with [Fe/H] data is more effective at identifying FP TOI targets with lower metallicity, which aligns with the established correlation between stellar metallicity and planet occurrence rates.
More specifically, the comparison yielded the following insights:
\begin{itemize}
    \item The accuracy against the TFOPWG results decreased from 76\% to 72\% when [Fe/H] was excluded as a training feature. This indicates that [Fe/H] plays a significant role in enhancing the model's overall accuracy.
    \item The number of TP TOIs identified by both models, with and without [Fe/H], remained largely similar. However, the number of TN TOIs decreased by 51 when [Fe/H] was not included. Consequently, the precision of the model dropped from 67.4\% to 62.4\%, suggesting that [Fe/H] contributes to the model's ability to correctly identify non-planetary targets.
    \item The proportion of FP predicted of the total TOI targets decreased from 38.6\% to 32.5\% when [Fe/H] was excluded. This implies that incorporating [Fe/H] as a feature aids the model in identifying more false positives.
\end{itemize}

\begin{figure}
    \centering
    \includegraphics[width=\textwidth]{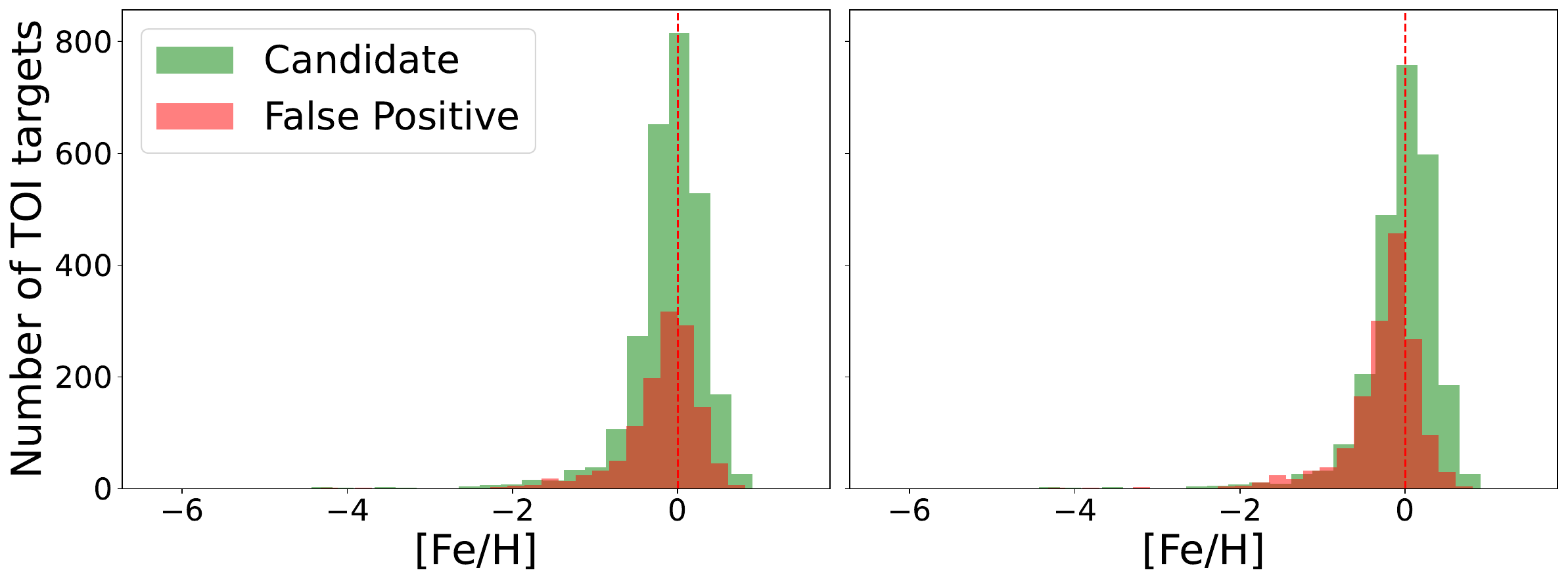}
    \caption{Histograms of Metallicity of Predicted Planets and Non-Planets of TOIs. \textit{Left:} Prediction by the Random Forest model trained without [Fe/H]. \textit{Right:} Prediction by the Random Forest model trained with [Fe/H]. The latter model finds more non-planets orbint stars with lower metallicity. The red dashed line indicates the position of [Fe/H] = 0. }
    \label{fg:metallicity_histogram}
\end{figure}

\section{Further Analysis on Planet Candidates}
\subsection{Habitable Exoplanet Candidates}
We further conducted an in-depth analysis of all 1595 TOI planet candidates predicted by the trained model but not yet confirmed by TFOPWG\footnote{PC and APC targets in Table \ref{tb:toi_prediction}.}, with the aim of searching for potential habitable worlds that reside within the habitable zones of their host stars.

The habitable zone is defined as the region around a star where conditions are right for water to exist in its liquid state. It is primarily determined by the stellar luminosity and effective temperature. Here, we estimated the boundaries ($d_\mathrm{HZ}$) of the habitable zone by following the formulas presented by \citet{2013ApJ...765..131K}:
\begin{equation}
    d_\mathrm{HZ} = \left(\frac{L_\star / L_\odot}{S_\mathrm{eff}}\right)^{0.5},
    \label{eq:hz}
\end{equation}
where $d_\mathrm{HZ}$ is in the unit of AU, and $S_\mathrm{eff}$ is the stellar fluxes reaching the top of the atmosphere of an Earth-like planet defined as:
\begin{equation}
    S_\mathrm{eff} = S_\mathrm{eff,\odot} + a T_\star + b T_\star^2 + c T_\star^3 + d T_\star^4,
    \label{eq:seff}
\end{equation}
where $T_\star = T_\mathrm{eff} - 5780$ K, and the solar fluxes $S_\mathrm{eff,\odot}$ as well as the coefficients $a$, $b$, $c$, and $d$ are given in Table~3 of \citet{2013ApJ...765..131K}. Consequently, we can determine the inner ($d_\mathrm{HZ, inner}$) and outer ($d_\mathrm{HZ, out}$) boundaries of the habitable zone for each planet candidate using the moist-greenhouse and maximum greenhouse limits provided in Table~3 of \citet{2013ApJ...765..131K}, respectively.

In order to determine whether a planet candidate is loacted within the habitable zone, we calculated the orbital semi-major axis $a_\mathrm{orb}$ for each planet candidate using Kepler's third law, assuming planet mass is negligible compared to the host star mass:
\begin{equation}
    a_\mathrm{orb} = \left(\frac{G M_\star P^2}{4 \pi^2}\right)^{1/3},
    \label{eq:kepler}
\end{equation}
where $G$ is the gravitational constant, $M_\star$ is the mass of the host
star that are obtained from Gaia DR3.
We then compared $a_\mathrm{orb}$ with $d_\mathrm{HZ, inner}$ and $d_\mathrm{HZ, out}$ to determine whether the planet candidate is located within the habitable zone.

In total we identified 20 TOI planet candidates with an $a_\mathrm{orb}$ that falls between $d_\mathrm{HZ, inner}$ and $d_\mathrm{HZ, out}$, which places them within the habitable zone of their respective host stars. And it was noticed that out of the 20 candidates, 4 planets have an $R_\mathrm{p}$ less than 2.5 (R$_\oplus$). \citet{2014ApJ...795...65J} studied several hydrodynamic escape models and found that 2.5 $R_\oplus$ may be a suitable threshold where planets transit from being rocky to having thick atmospheres (gaseous planets). 
In Table~\ref{tb:habitable_planets}, we present the key parameters of these planets as well as their host stars.

\begin{sidewaystable}[htbp]
    \caption{Table of Planets residing in the habitable zone of their host stars. The TOI column indicates the TOI number of the planets with their properties listed in the last three columns. The rest of the columns are the stellar parameters of the host stars. The last five rows are objects in multi-planet systems.}
    \label{tb:habitable_planets}
    \centering
    \small
    \begin{tabular}{lcccccccccccc}
        \toprule
        TIC  & TOI & $D$ & $R_{*}$ & $T_{\text{eff}}$ & [Fe/H] & $L_\star$ & $M_\star$ & $d_\mathrm{HZ, in}$ & $d_\mathrm{HZ, out}$ & $P$ & $R_{p}$ & $a_\mathrm{orb}$ \\
        & & (PC) & $(R_{\odot})$ & (K) & & ($L_\odot$) & ($M_\odot$) & (AU) & (AU) & (d) & $(R_{\oplus})$ & (AU) \\
        \hline
 323295479 & 1861.01 & 134.62 & 1.061 & 5490 & 0.350 & 0.957 & 0.888 & 0.983 & 1.711 & 742.51 & 11.19 & 1.543 \\
 384159646 & 1895.01 & 205.12 & 1.374 & 6762 & -1.211 & 2.927 & 1.239 & 1.638 & 2.705 & 748.07 & 9.57 & 1.732 \\
 167905035 & 4191.01 & 83.57 & 1.102 & 5816 & -0.370 & 1.247 & 0.948 & 1.107 & 1.898 & 742.86 & 2.28 & 1.577 \\
 146282666 & 4298.01 & 124.27 & 1.221 & 5857 & -0.820 & 1.907 & 1.092 & 1.367 & 2.339 & 722.26 & 2.72 & 1.622 \\
 334632624 & 4361.01 & 98.09 & 1.220 & 5944 & -0.130 & 1.666 & 1.007 & 1.274 & 2.171 & 741.43 & 2.77 & 1.607 \\
 304339227 & 4629.01 & 91.31 & 1.126 & 6324 & -0.270 & 1.474 & 1.028 & 1.180 & 1.981 & 714.70 & 2.49 & 1.579 \\
 415399713 & 5527.01 & 144.37 & 0.926 & 5862 & 0.081 & 0.760 & 0.928 & 0.863 & 1.476 & 346.16 & 2.54 & 0.941 \\
 88565745 & 5571.01 & 307.78 & 1.100 & 6250 & -0.625 & 1.638 & 1.099 & 1.247 & 2.099 & 731.48 & 11.78 & 1.639 \\
 85292615 & 5721.01 & 138.49 & 1.007 & 5881 & -0.130 & 1.025 & 1.008 & 1.001 & 1.711 & 729.48 & 3.33 & 1.590 \\
 162631539 & 5724.01 & 50.36 & 1.375 & 6061 & -0.560 & 2.316 & 1.143 & 1.495 & 2.535 & 697.40 & 3.02 & 1.609 \\
 29385858 & 5990.01 & 102.07 & 1.000 & 5626 & -0.090 & 0.890 & 0.906 & 0.943 & 1.631 & 700.96 & 2.22 & 1.494 \\
 296737508 & 6267.01 & 117.75 & 1.270 & 5700 & 0.030 & 1.434 & 0.986 & 1.193 & 2.056 & 733.99 & 4.81 & 1.585 \\
 313671132 & 6559.01 & 175.55 & 1.220 & 5878 & 0.060 & 1.502 & 0.980 & 1.212 & 2.073 & 733.38 & 6.17 & 1.581 \\
 192014454 & 6665.01 & 109.74 & 1.245 & 5973 & -0.320 & 1.714 & 1.054 & 1.290 & 2.197 & 735.80 & 2.77 & 1.623 \\
 206361691 & 6668.01 & 99.24 & 1.395 & 6002 & -0.260 & 2.290 & 1.068 & 1.490 & 2.533 & 718.65 & 3.45 & 1.605 \\
 232608943 & 4600.02 & 215.95 & 0.832 & 5073 & 0.330 & 0.401 & 0.808 & 0.647 & 1.148 & 482.82 & 6.91 & 1.122 \\
 149989864 & 6690.01 & 158.66 & 1.068 & 5599 & 0.020 & 0.933 & 0.915 & 0.966 & 1.673 & 761.71 & 2.56 & 1.585 \\
 149302744 & 699.03 & 192.86 & 1.419 & 6139 & 0.040 & 2.370 & 1.179 & 1.507 & 2.548 & 672.64 & 2.97 & 1.587 \\
 469810663 & 7060.02 & 190.56 & 1.350 & 6010 & -0.180 & 1.878 & 1.119 & 1.348 & 2.292 & 1064.70 & 4.45 & 2.119 \\
 306472057 & 791.02 & 333.96 & 1.430 & 6572 & -1.623 & 3.711 & 1.336 & 1.856 & 3.085 & 928.05 & 12.07 & 2.051 \\        
 \bottomrule
    \end{tabular}
\end{sidewaystable} 

Additionally, we found that five habitable zone planets are part of multi-planet systems.
TIC~232608943 has two giant planets, one is close to the host star at 0.346 AU, while the outer one is at 1.122 AU and is within the habitable zone. On contrast, TIC~149989864 has two mini-Neptune sized planets with the inner one just inside the outer edge of the habitable zone. TIC~469810663 has a very close orbitor at 0.066 AU and a second planet at 2.119 AU within the habitable zone. TIC~306472057 is a system with two gas--giant planets, one of which is located at 0.570 AU from its host star, while the other is located at 2.051 AU and is within the habitable zone. 
TIC~149302744 is a triple-planetary system with three super-Earth to mini-Neptune sized planets. The closest two planets are located within 0.22 AU from the host star, while the third planet is located at 1.587 AU from the same star and is within the habitable zone. Although these five habitable zone planets are not Earth-sized or unlikely to be rocky, they are still worth further follow-up observations and studies, as their satellite moons could potentially be habitable \citep[e.g.,][]{2018ApJ...860...67H}. The stellar parameters of the host stars in these multi-planetary systems, along with those of the Sun for comparison, are presented in the last five rows of Table~\ref{tb:habitable_planets}. The planetary properties, including those of the terrestrial planets in the solar system for comparison, are listed in Table~\ref{tb:habitable_planets_in_multi_system}.

\begin{table}[h]
    \caption{Planetary properties of the planets in the multi-planet systems with habitable zone planets. The planets in the solar system are also included for comparison.}
    \label{tb:habitable_planets_in_multi_system}
    \centering
    \begin{threeparttable}
        \begin{tabular}{llccc}
            \toprule
            Host & Planet/TOI & $P$ (d) & $R_\mathrm{p}$ (R$_\oplus$) & $a_\mathrm{orb}$ (AU) \\
            \midrule
            Sun & \small{Mercury} & 87.874 & 0.383 & 0.387 \\
            & Venus & 224.700 & 0.950 & 0.723 \\
            & Earth & 365.256 & 1.000 & 1.000 \\
            & Mars & 687.000 & 0.532 & 1.524 \\
            \midrule
            TIC~232608943 & 4600.01 & 82.690 & 6.824 & 0.346 \\
            & $4600.02^*$ & 482.819 & 6.909 & 1.122 \\
            \midrule            
            TIC~149989864 & $6690.01^*$ & 761.709 & 2.558 & 1.585 \\
            & 6690.02 & 904.535 & 2.731 & 1.777 \\
            \midrule          
            TIC~149302744 & 699.01 & 14.801 & 2.448 & 0.125 \\
            & 699.02 & 33.634 & 2.305 & 0.215 \\
            & $699.03^*$ & 672.637 & 2.975 & 1.587 \\
            \midrule
            TIC~469810663 & 7060.01 & 5.850 & 2.713 & 0.066 \\
            & $7060.02^*$ & 1064.700 & 4.452 & 2.119 \\
            \midrule            
            TIC~306472057 & 791.01 & 139.305 & 10.467 & 0.579 \\
            & $791.02^*$ & 928.046 & 12.067 & 2.051 \\
            \bottomrule
        \end{tabular}
        \begin{tablenotes}
            \scriptsize
            \item \textbf{Notes.} $^*$ indicates a planet within the habitable zone.
        \end{tablenotes}
    \end{threeparttable}
\end{table}   

The discovery of these habitable zone planets underscores the potential of the ML model in identifying promising exoplanet candidates. These planets are excellent candidates for follow-up observations to confirm their existence and further investigate their properties, including their atmospheric composition and potential habitability.

\subsection{Interesting TESS targets for follow-up observation}

In addition to the 20 hosts of habitable zone planets, we further analyzed the TOI planet candidates predicted by the trained model, with the intention of finding planet candidates for follow-up observations. Generally, we focused on planets with $P$ less than 30 days, distance $D$ less than 150 pc, and bright enough with a Gaia G-band magnitude $G_\mathrm{mag} < 11$, which make them more suitable for ground-based follow-up observations. A short orbital period allows for multiple orbits to be detected within a limited time, and a close distance makes it easier to be detected.
\begin{table}[h!]
    \caption{List of proposed planet candidates for follow-up observations. The first 4 candidates are gas giants, while the rest 4 are Earth-sized to mini-Neptunes.}
    \label{tb:follow-up_candidates}
    \centering
    \resizebox{\textwidth}{!}{
        \begin{tabular}{l c c c c c c c c}
        \toprule
        TOI & TIC & RA & Dec & $P$ & $D$ & $R_\mathrm{p}$ & $G_\mathrm{mag}$  & [Fe/H] \\
        & & (deg) & (deg) & (d) & (pc) & (R$_\oplus$) & & \\
        \hline
        1533.02 & 345143460 & 355.247 & 57.485 & 8.064 & 99.76 & 9.29 & 10.68 & 0.36 \\
        173.01 & 270341214 & 33.459 & -80.583 & 29.754 & 149.86 & 11.99 & 9.2 & -0.17 \\
        4492.01 & 298297931 & 308.651 & 29.238 & 4.433 & 142.33 & 12.52 & 10.09 & 0.17 \\
        5394.01 & 61109252 & 154.575 & 17.396 & 15.194 & 64 & 9.68 & 8.41 & -0.356 \\
        \midrule
        5140.01 & 175193677 & 129.689 & 23.685 & 15.611 & 35.2 & 2.9 & 6.76 & -0.076 \\
        554.01 & 407966340 & 60.748 & 9.208 & 7.049 & 45.11 & 2.85 & 6.79 & -0.32 \\
        6965.01 & 80224448 & 103.678 & 24.245 & 5.969 & 31.02 & 1.35 & 6.72 & -0.24 \\
        6965.02 & 80224448 & 103.678 & 24.245 & 28.069 & 31.02 & 1.62 & 6.72 & -0.24 \\
        \bottomrule
        \end{tabular}
    }
\end{table}

We obtained the stellar distance $D$ from \citet{2021AJ....161..147B} and $G_\mathrm{mag}$ from Gaia DR3. Figures~\ref{fg:toi_radius_orbperiod}, \ref{fg:toi_radius_rmedgeo}, and \ref{fg:toi_radius_gmag} show the distribution of $P$, $D$, and $G_\mathrm{mag}$ in different combinations. Based on these distributions, we selected eight interesting planets for follow-up observations, as shown in Table~\ref{tb:follow-up_candidates}. Among them, four are giant planets, including one multiplanet system, while the other four range from Earth-sized to mini-Neptunes, with the smallest having a radius of 1.35 R$_\oplus$. These targets are promising candidates for further study and could provide valuable insights into the diversity of exoplanets.

\begin{figure}
    \centering
    \includegraphics[width=\textwidth]{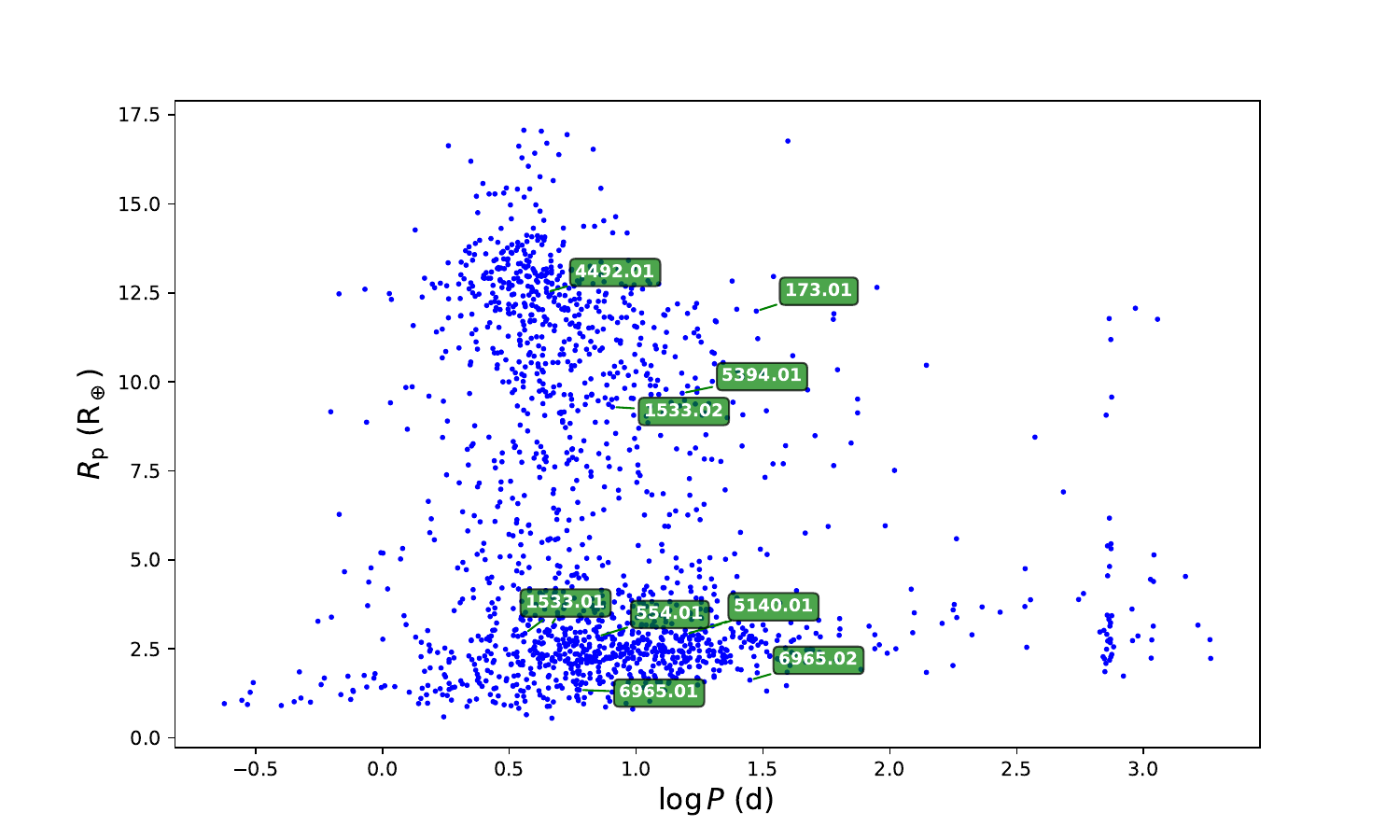}
    \caption{TOI radius vs orbital period. The TOI numbers of the proposed planet candidates for follow-up observations, listed in Table~\ref{tb:follow-up_candidates}, are highlighted in green. Additionally, TOI-1533.01 is also highlighted as the companion of TOI-1533.02.}
    \label{fg:toi_radius_orbperiod}
\end{figure}

\begin{figure}
    \centering
    \includegraphics[width=\textwidth]{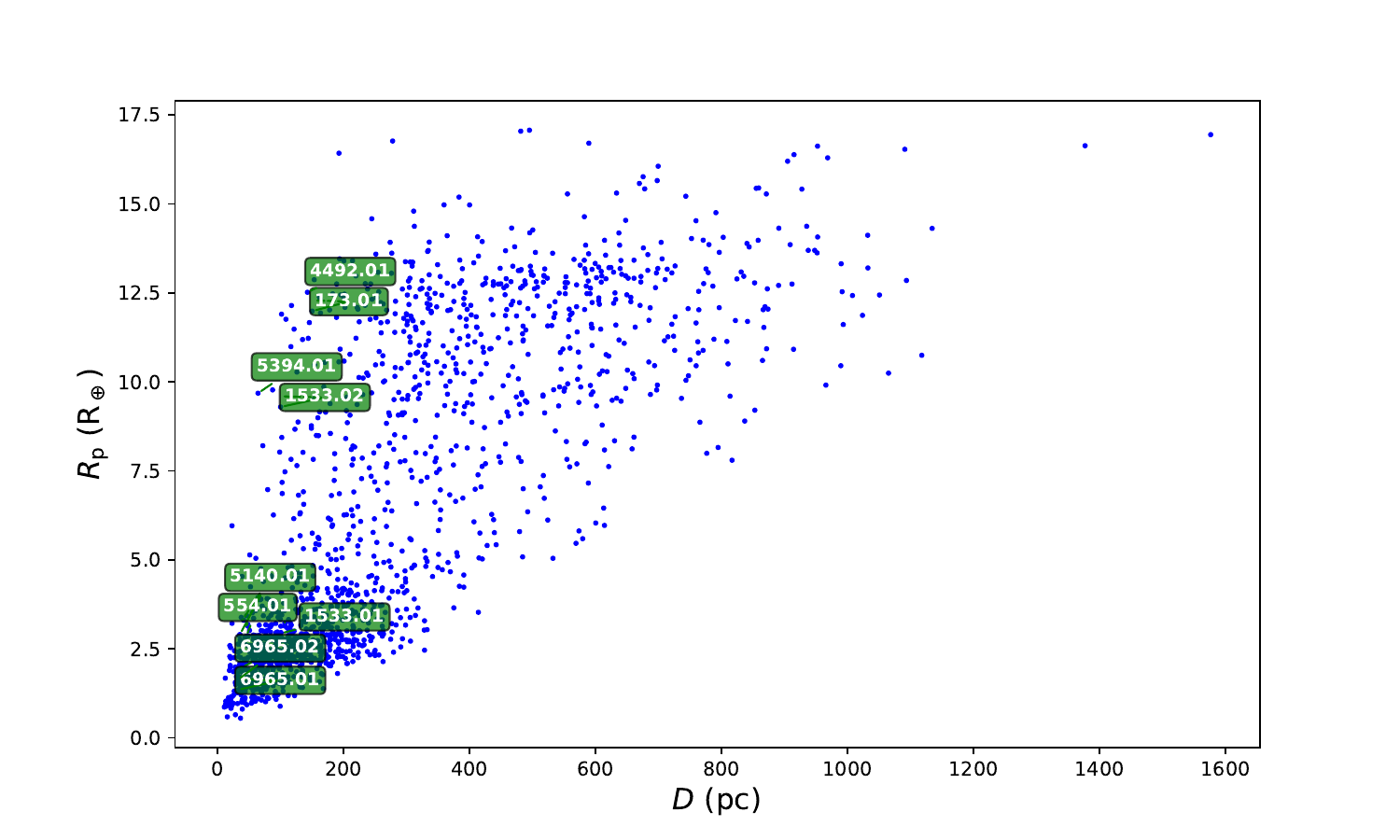}
    \caption{TOI radius vs stellar distance. The symbol notation is the same as in Figure~\ref{fg:toi_radius_orbperiod}.}
    \label{fg:toi_radius_rmedgeo}
\end{figure}

\begin{figure}
    \centering
    \includegraphics[width=\textwidth]{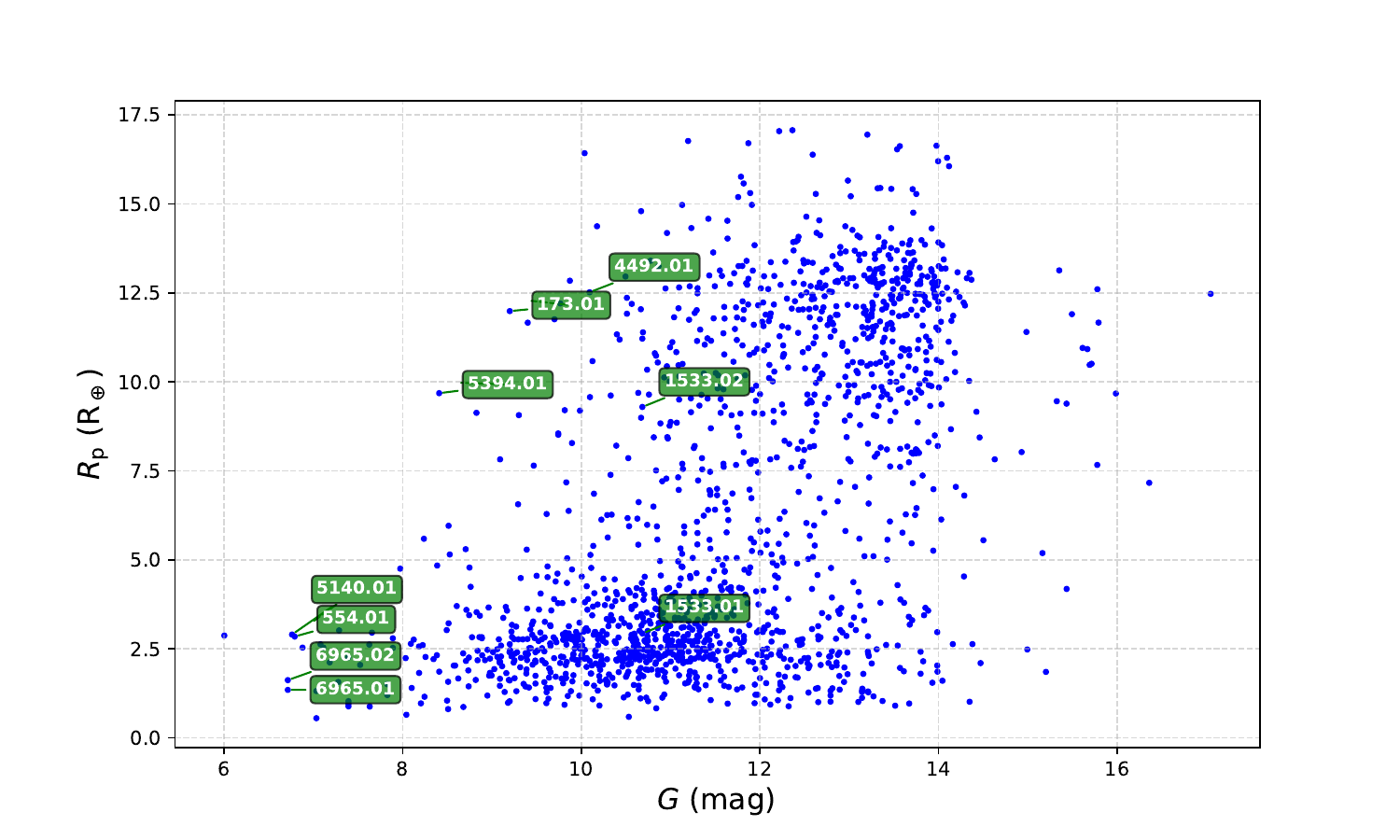}
    \caption{TOI radius vs Gaia G band magnitude. The symbol notation is the same as in Figure~\ref{fg:toi_radius_orbperiod}.}
    \label{fg:toi_radius_gmag}
\end{figure}

\subsection{Analysis on TESS Planet Candidate Population}
\label{sc:tess_population}
We further conducted a statistical analysis to compare the population of TOI planet candidates with that of the Kepler planet candidates. A total of 4502 KOIs dispositioned as confirmed planet or candidates were selected for comparison. For TOIs, the sample consists of two parts: (1) TOIs catagroized by TFOPWG dispositions as CP (confirmed planet) and KP (known planet), and (2) TOIs predicted as planet candidates by our model and catagorized by TFOPWG dispositions as PC (planetary candidate) and APC (ambiguous planetary candidate). Thus, in total, 2514 TOIs were selected for comparison.
We compared the two samples of TOI and KOI targets in terms of $R_\mathrm{p}$ and $P$, host star spectral types, and multi-planete systems. 

\subsubsection{Planet Radius and Orbital Period}

\begin{figure}
    \centering
    \includegraphics[width=\textwidth]{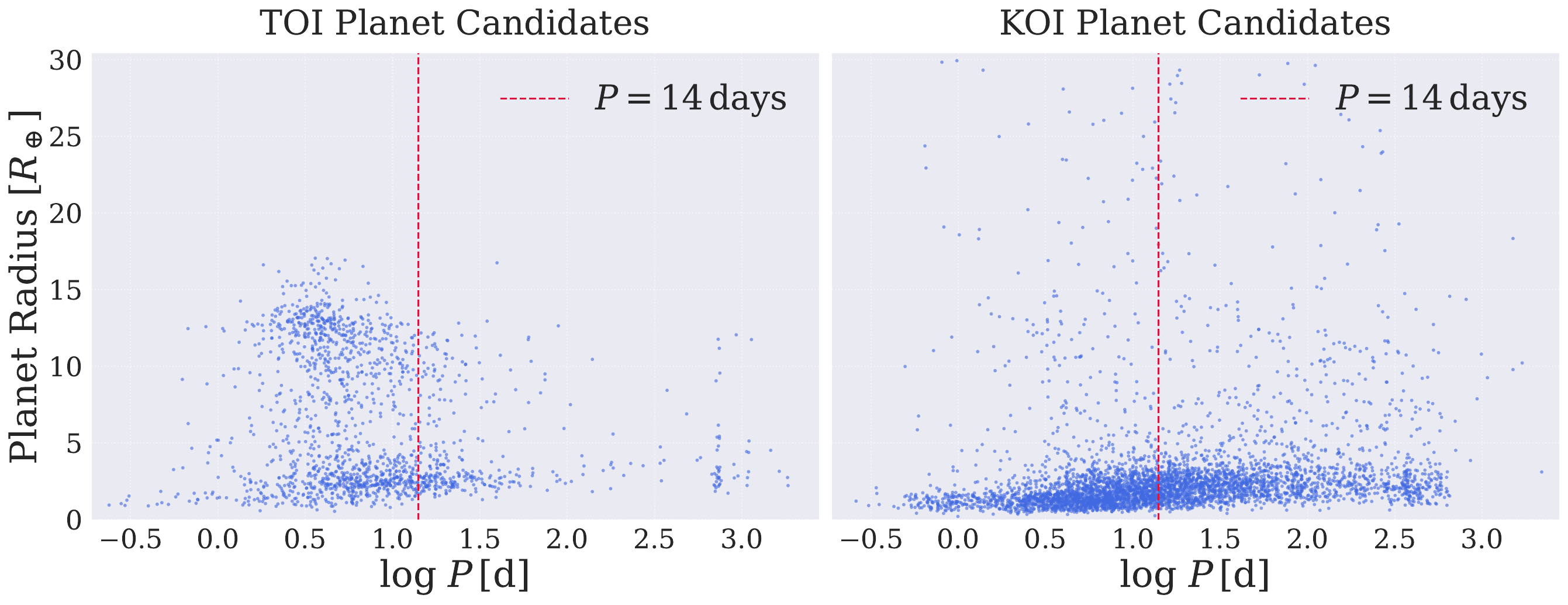}
    \caption{Side-by-side comparison of TOI (left) and KOI (right) radius vs orbital period. The red dashed line marks the 14-day orbital period. TOI detected more giant planets with shorter orbital periods, while KOI detected more smaller planets with longer orbital periods.}
    \label{fg:toi_koi_radius_orbperiod}
\end{figure}

Figure~\ref{fg:toi_koi_radius_orbperiod} illustrates a side-by-side comparison of the planet candidates from both TOI and KOI, plotted in terms of $R_\mathrm{p}$ against $P$. Both samples highlight a concentration of planet candidates within the radius range of 0.25--4 R$_\oplus$, corresponding to Earth-like to Neptune-like planets.
However, TOI has a prominent cluster of hot Jupiters with radii around 10-15 R$_\oplus$ and orbital periods less than 14 days, accounting for 28.5\% of the total TOI planet population. In contrast, such hot Jupiters only account for 1.3\% of the total KOI planet population. This discrepancy can be attributed to the different observation strategies of the TESS and Kepler missions. TESS is an all-sky survey and observes most of the targets only for a few sectors, each lasting 28 days, favoring the detection of planets with short orbital periods. On the other hand, Kepler observed the same field of view for over 9 years, allowing for the detection of planets with longer orbital periods. 

\subsubsection{Host Star Spectral Type}

\begin{figure}
    \centering
    \includegraphics[width=\textwidth]{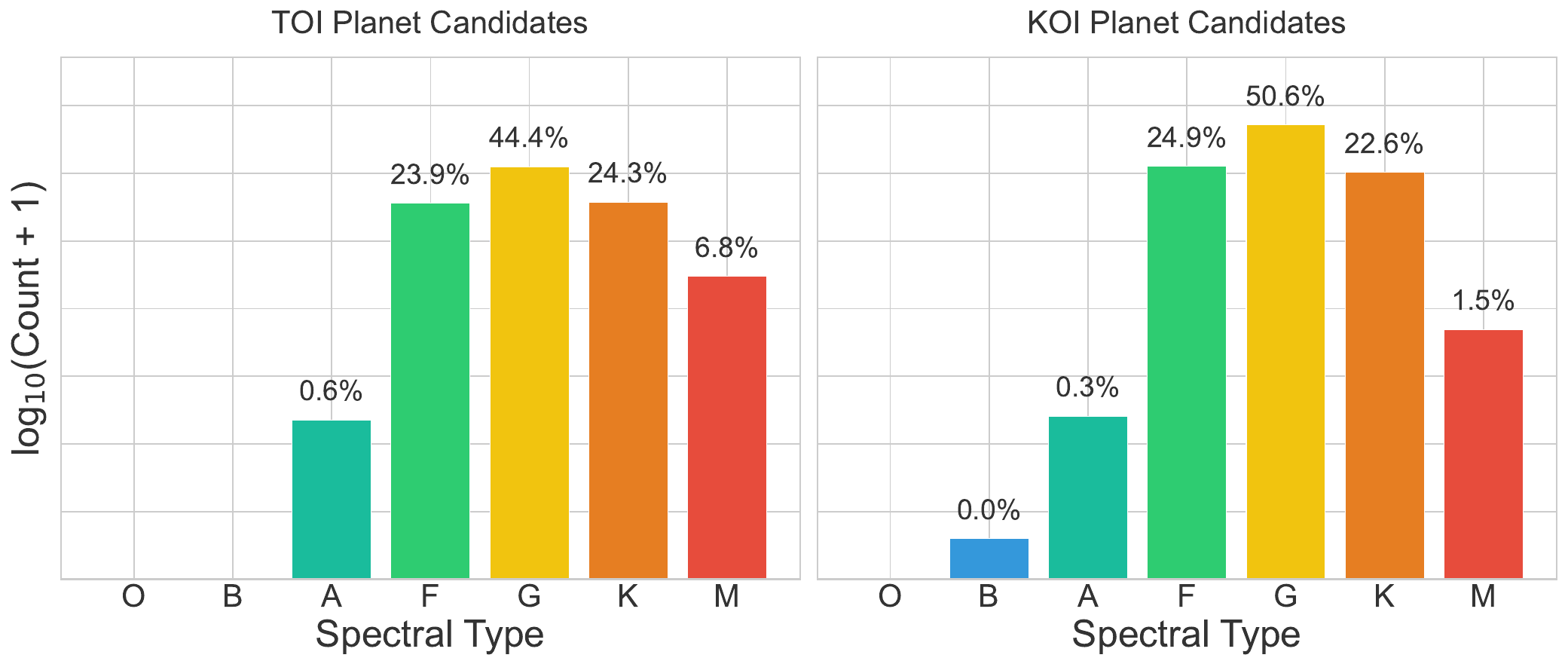}
    \caption{Side-by-side comparison of TOI (left) and KOI (right) host stars' spectral types. The TOI table contains more M-type stars, but missing B-type stars.}
    \label{fg:toi_koi_spectral_type}
\end{figure}

The distinct observational strategies between TESS and Kepler are further reflected in the spectral distribution of the host stars.
As shown in Figure~\ref{fg:toi_koi_spectral_type}, TESS demonstrates a stronger capability in detecting planets around M-type stars, which are abundant and have favorable conditions for transit detection, resulting in a significantly higher number of M-type stars compared to the KOI. Conversely, the absence of B-type stars in TOI highlights the challenges in detecting planets around these hot, luminous stars due to their bright background and rarity. In contrast, the Kepler mission, showcases a more even distribution across spectral types, including a notable presence of B-type stars, indicating that its longer observation period and targeted field of view were better suited for detecting planets around a wider range of stellar types.

A comprehensive investigation of the distribution of $G_{\mathrm{mag}}$ and $D$ for M-type stars hosting planets reveals significant differences in their observed populations. Over 92\% of M-type stars in the Kepler mission have $G_{\mathrm{mag}}$ values between 14 and 20, whereas 75\% of M-type stars in the TESS mission fall within the 8-14 range, representing a brighter population. Additionally, 80.8\% of M-type stars hosting TOI planet candidates are within 100 PC, while 80.6\% of Kepler M-type hosts are located between 200 and 800 PC.
These differences suggest that the lower number of detected planets around M-type stars by Kepler may be due to their significantly greater distances (hundreds to thousands of PC) compared to TESS’s closer targets (within 100 PC). The dim nature of M-type stars makes detecting transiting exoplanets increasingly challenging as the distance to the system increases, which is the case for Kepler. This analysis highlights the complementary roles of the two missions, with TESS focusing on nearby, brighter stars, while Kepler targeted more distant, fainter ones.

\subsubsection{Multi-Planet Systems}

\begin{figure}
    \centering
    \includegraphics[width=\textwidth]{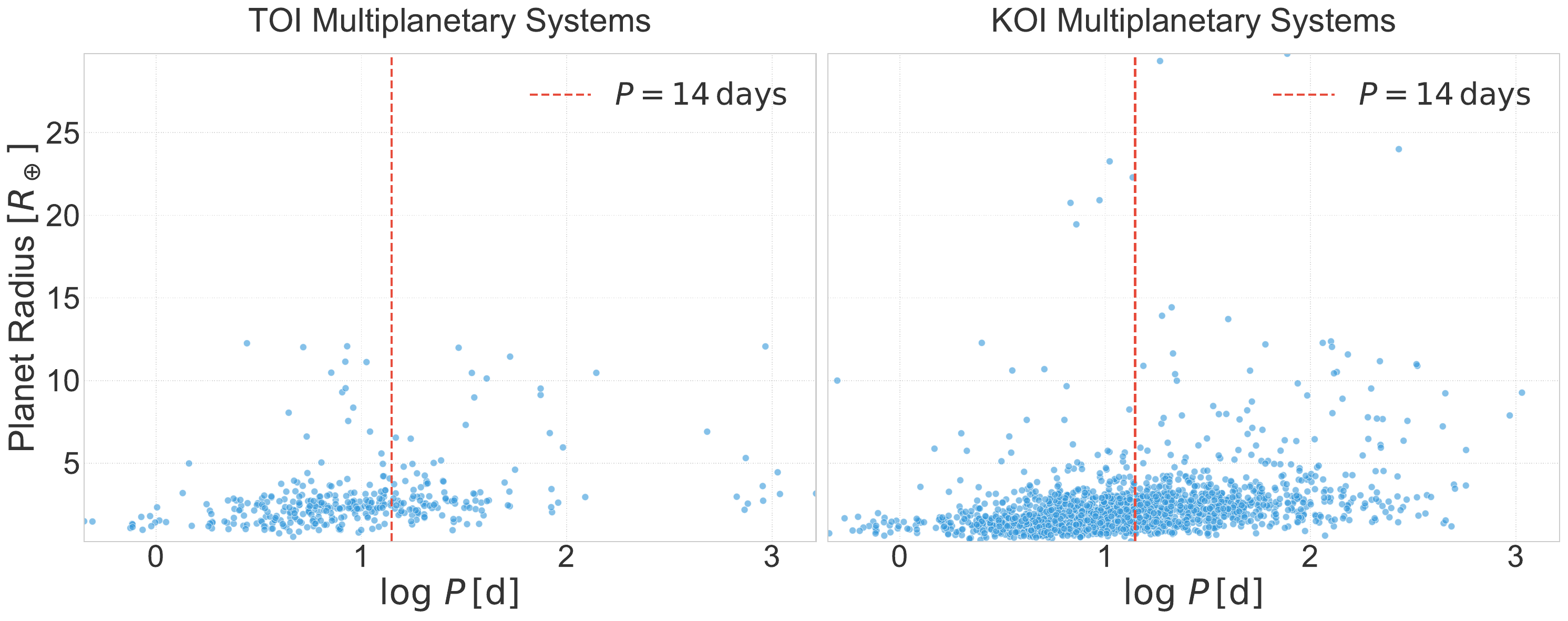}
    \caption{Side-by-side comparison of TOI (left) and KOI (right) radius vs orbital period for multi-planet systems. The red dashed line marks the 14-day orbital period. Compared with Figure~\ref{fg:toi_koi_radius_orbperiod}, short-period giant planets are absent in this figure, suggesting that they tend to exist as single-planet systems.}
    \label{fg:toi_koi_radius_orbperiod_multipl}
\end{figure} 

\begin{table}[ht]
    \caption{Number of multi-planet systems - TOI vs KOI. The percentage of multi-planet systems out of total planet candidates identified is also included in the last row.}
    \label{tb:multiplanetary_systems}
    \centering
    \begin{tabular}{l c c c c c c}
        \toprule
        & \multicolumn{2}{c}{TESS} & & \multicolumn{2}{c}{Kepler} \\
        \cmidrule{2-3} \cmidrule{5-6}
        & Hosts & Planets & &Hosts & Planets \\
        \midrule
        Multi-planet system & 166 & 379 & & 712 & 1784 \\
        Total planet candidates identified & 2301 & 2514 & & 3430 & 4502 \\
        \midrule
        Percentage & 7.2\% & 15.1\% & & 20.8\% & 39.6\% \\
        \bottomrule
    \end{tabular}
\end{table}

As shown in Table~\ref{tb:multiplanetary_systems}, 15.1\% of all predicted TESS planet candidates by our model are part of multi-planet systems, with 7.2\% of host stars hosting multiple planets. In contrast, Kepler's multi-planet systems account for 39.6\% of all planet candidates, with 20.8\% of host stars hosting multiple planets. This significant difference between TESS and Kepler can be attributed again to the different observation strategies of the two missions. TESS's observation sectors last only 28 days, favoring the detection of planets with short orbital periods, while planets with longer orbital periods in the same planetary system may not be detected. Additionally, compared with Figure~\ref{fg:toi_koi_radius_orbperiod} where a cluster of hot Jupiters are identified in the TOI sample, Figure~\ref{fg:toi_koi_radius_orbperiod_multipl} shows that the cluster is significantly reduced when considering only multi-planet systems. Indeed, only 0.8\% of Jupiter-sized planets discovered by TESS are part of multi-planet systems, supporting the hypothesis of high-eccentricity tidal migration as a probable mechanism for the formation of these hot Jupiters \citep[for a review see,][]{2018ARA&A..56..175D}. During this migration process, the giant planet’s orbit becomes highly elongated. As it moves inward, it can gravitationally disturb, scatter, or eject other planets. This often leads to the destruction or removal of nearby companions. As a result, hot Jupiters typically end up as the only close-in planet in their system \citep{2015ApJ...808...14M}.

\section{Discussion}
It was surprising to observe that, despite the overall comparable performance of the Random Forest model and the Transformer neural network when validating with the KOI data, their prediction accuracies diverged significantly when applied to TOI data. Specifically, the Random Forest model outperformed the Transformer model, particularly in identifying confirmed planets. As initially analyzed in Section \ref{sc:toi_prediction}, this discrepancy may be attributed to the substantial differences in the distribution patterns of KOI and TOI data across several key features, such as orbital period ($P$). Additionally, the limited size of the training dataset available in the KOI table may have constrained the Transformer model's ability to generalize effectively.

One of the most significant challenges encountered in this work was the lack of complete and accurate data in the TOI table. As shown in Section~\ref{sc:tess_data_prep}, only 72.7\% of the TOIs could be found in Gaia DR3 by matching the identifiers of the targets in SIMBAD. Furthermore, an additional 9.8\% of TOI targets had to be excluded due to the fact that metallicity data or the necessary parameters required to convert metallicity from [M/H] to [Fe/H] are unavailable in Gaia DR3. This left only 62.9\% of the total TOI targets available for prediction.

The fragmented and incomplete nature of the available data posed significant challenges for model training and final prediction. Had the mapping between important astronomy catalogs been more comprehensive and had more observations been conducted to measure metallicity more accurately, the analysis would have been less time-consuming, and the trained model could have achieved greater accuracy. The current state of data integration and availability underscores the need for enhanced database interoperability and more extensive observational campaigns.

On a positive note, the process of joining data from different astronomy databases, such as Gaia DR3 and the NASA Exoplanet Archive, as demonstrated in Section~\ref{sc:toi}, proved to be a highly beneficial approach. This integration not only enriched the dataset but also facilitated a more thorough and accurate analysis of the exoplanet candidates. Moving forward, the continued development of robust data integration tools and the expansion of observational datasets will be crucial in advancing the field of exoplanetary science and improving the reliability of predictive models.

\section{Conclusion}

In this study, four machine learning algorithms and a transformer-based deep learning model were trained and tested using transit data and stellar features obtained from the Kepler Mission. The evaluation of these models revealed that all of them performed reasonably well, with three of them achieving accuracies exceeding 80\%. 
The Random Forest model outperforms all other algorithms assessed, achieving a precision of 0.839, recall of 0.852, accuracy of 0.839, and an F1 score of 0.846. These results reflect the strong capability of the model in correctly identifying true positives while maintaining high overall accuracy. Additionally, its consistent performance across multiple metrics underscores its effectiveness and reliability in classifying exoplanet candidates.

The Transformer neural network model ranks second in terms of performance, with an accuracy of 0.826 and an F1 score of 0.837, narrowly trailing the Random Forest model. Despite its current slight underperformance, the Transformer model holds considerable promise for future improvement. The limited size of our training dataset, which consists of about 7000 targets, currently restricts the model's capacity. Nevertheless, as the dataset grows to encompass a wider variety of exoplanets with diverse characteristics, the Transformer model is anticipated to exhibit substantial performance gains, leveraging its capability to identify intricate patterns and relationships.

When applied to 3987 TESS candidates, the best-performing Random Forest model identified 1595 high-confidence new planet candidates and correctly recovered 86\% of all previously confirmed TESS exoplanets in a blinded validation test. A in-depth analysis of these new candidates reveals that 100 previously unrecognized multi-planet systems

Our analysis of multi-planetary systems suggests that short-period giant planets are often absent in such configurations, reinforcing the idea that hot Jupiters are more commonly found in isolation. This has implications for planetary formation and migration theories, particularly regarding the dynamical interactions that may lead to the ejection or inhibition of additional planets in systems hosting a hot Jupiter.

Furthermore, our study also identifies 20 planet candidates residing in the habitable zone of their host stars, offering promising targets for future follow-up observations. These planets, which could potentially support liquid water, are of great interest for studying planetary habitability and the conditions necessary for life. Their detection and characterization will be crucial for expanding our understanding of habitable worlds beyond our Solar System.

For future observations, we have shortlisted eight planet candidates for follow-up, selected based on their location in the habitable zone, planetary size, and host star characteristics. This list includes four giant planets and four smaller planets, ranging from Earth-sized to mini-Neptunes, with the smallest having a radius of 1.35 R$_\oplus$. Observational strategies will involve spectroscopic analysis to determine atmospheric composition, transit timing variations to refine orbital properties, and potential direct imaging efforts for the most promising candidates. These follow-up observations will be conducted using a combination of ground-based telescopes and space missions, such as JWST and future high-resolution spectrographs, to further investigate the atmospheric and physical properties of these planets, and upcoming PLATO and Earth 2.0 missions to search for Earth-like planets.

Our study not only evaluates the predictive capabilities of different classification models but also provides insights into exoplanet population distributions. Future work will focus on refining the Transformer-based approach with larger datasets and exploring additional astrophysical parameters to further enhance classification accuracy. Furthermore, integrating additional data sources and leveraging advanced deep-learning techniques may yield even more reliable and interpretable results, ultimately improving our understanding of exoplanetary systems.

\end{document}